\documentclass[reprint, amsmath, amssymb, aps, prb, superscriptaddress]{revtex4-2}
\usepackage{tabularx}
\usepackage{graphicx}
\usepackage{dcolumn}
\usepackage{bm, color}
\usepackage{amsmath}
\usepackage{amsfonts}
\usepackage{amssymb}
\usepackage{multirow}
\providecommand{\U}[1]{\protect\rule{.1in}{.1in}}

\newcommand{\vv}[1]{\boldsymbol #1}
\newcommand{\ket}[1]{\vert{#1}\hspace{0.45pt}\rangle}

\usepackage{hyperref}
\hypersetup{
	colorlinks = true,
}
\usepackage{physics}

\begin{document}

\title{Magnetic response of Majorana Kramers pairs with an order-two symmetry}

\author{Yuki Yamazaki}
\affiliation{Department of Physics, Nagoya University, Nagoya 464-8602, Japan}
\author{Shingo Kobayashi}
\affiliation{RIKEN Center for Emergent Matter Science, Wako, Saitama 351-0198, Japan}
\author{Ai Yamakage}
\affiliation{Department of Physics, Nagoya University, Nagoya 464-8602, Japan}

\date{\today}

\begin{abstract}
We study an intrinsic relation between the topology of bulk electronic states and magnetic responses of Majorana Kramers pairs, Kramers pairs of Majorana fermions, on a surface of time-reversal-invariant topological superconductors. Majorana Kramers pairs respond to an applied magnetic field anisotropically due to the interplay between time-reversal and crystalline symmetries. In this paper, we propose a systematic procedure to determine such surface magnetic responses in systems with an order-two symmetry. From the analysis of topological invariants associated with an order-two symmetry, it is found that magnetic responses are classified into four types, which are attributed to different topological invariants and exhibit distinguishable, characteristic magnetic responses. 
For a Kramers pair of Majorana fermions protected by $\mathbb{Z}_2$ topological invariants, we clarify that types of magnetic responses are determined only from Fermi--surface topology and symmetry of Cooper pairs.
Finally, we apply our theory to the topological nonsymmorphic crystalline superconducting state in UCoGe, which exhibits a biaxially anisotropic magnetic response. 
\end{abstract}

\maketitle

\makeatletter
\def\ext@table{}
\makeatother
\makeatletter
\def\ext@figure{}
\makeatother

\section{Introduction} 
In the last decade, the study of topological superconductors (TSCs) has been a fascinating subject of unconventional superconductors, since they host emergent Majorana fermions on their surface as surface zero-energy Andreev bound states~\cite{Hu1526, Kashiwaya1641, Hasan3045, Qi1057, Tanaka011013, Sato076501, Haim2019a}. The emergent Majorana fermions follow non-Abelian statistics and are immune to a local noise as long as the superconducting gap remains in the bulk. These peculiar properties make TSCs a potential platform for fault-tolerant topological quantum computation \cite{Nayak1083}. 

A lot of effort to search for emergent  Majorana fermions has been devoted to a large variety of systems, such as proximity-induced superconductivity in nanowire~\cite{Sato020401, Kitaev131, Lutchyn077001, Oreg177002, Cook201105, Alicea076501} and magnetic atomic chains \cite{Choy195442, Nadj-Perge020407, Klinovaja186805, Vazifeh206802, Nadj-Perge602, Ruby197204}, and intrinsic superconductivity in doped topological (crystalline) insulators~\cite{Hor057001,Fu097001,Sasaki217001,Sasaki217004,Hashimoto174527,Fu100509,Matano852,Yonezawa123} and Dirac semimetals~\cite{Aggarwal3237,Wang3842,Kobayashi187001,Hashimoto014510,Oudah13617,Kawakami041026}. While the former systems break time-reversal symmetry (TRS), the latter does not. Following the Altland-Zirnbauer (AZ) classification~\cite{Altland1142, schnyder, kitaev, ryu}, TSCs with (without) TRS belong to class DIII (D), which hosts a Kramers pair of Majorana fermions (a Majorana fermion) on surfaces.
The difference relating to TRS is crucial for magnetic responses. 
Usually, a Majorana fermion is quite stable against any perturbation including magnetic fields due to charge neutrality, whereas Majorana Kramers pairs react to an applied magnetic field and exhibit a completely anisotropic Ising-like magnetic response due to the Kramers degeneracy and the self-conjugate property of Majorana fermions~\cite{Sato094504, shindou10, Chung235301, Nagato123603,  Mizushima12, Mizushima022001}. Another interesting effect is that the anisotropic behavior under a magnetic field can also induces a higher-order topological superconducting phase \cite{volpez19, plekhanov21}. 

There is currently intensive effort searching for topological phases in the presence of crystalline symmetry, such as reflection~\cite{ueno13,chiu-yao-ryu, morimoto-furusaki}, all order-two~\cite{shiozaki14}, nonsymmorphic~\cite{Shiozaki195413}, and rotational symmetries~\cite{Benalcazar224503, Fang01944}, point group~\cite{Cornfeld075105}, and magnetic point group symmetries~\cite{Shiozaki09354}. Those topological phases are called topological crystalline superconductors (TCSCs). Along these lines, Shiozaki and Sato have unveiled an interplay between the anisotropic magnetic response and crystalline symmetry through crystalline-symmetry-protected topological invariants~\cite{shiozaki14}.

In the previous studies~\cite{xiong17,kobayashi}, we have clarified that the $\mathbb Z$-invariant-protected Majorana Kramers pairs show anisotropic magnetic responses as a magnetic dipole (Ising) and octupole. In particular, a higher-rank magnetic response is intriguing, since it appears only in TCSCs with $J = 3/2$ fermions~\cite{kobayashi}.
We have also found a similar anisotropic magnetic response for $\mathbb{Z}_2$-invariant-protected Majorana Kramers pairs, along with a new type of quadrupole magnetic response for double Majorana Kramers pairs in TCSCs with nonsymmorphic symmetry~\cite{yamazaki07939}. 
Interestingly, those surface magnetic responses show a one-to-one correspondence to irreducible representations (irreps) of pair potentials under crystalline symmetry, which provides an experimental means to determine Cooper-pair symmetry in TSCs/TCSCs through magnetic responses of Majorana Kramers pairs.

In this paper, we elucidate a relation between $\mathbb{Z}_2$ topological invariants, anisotropic magnetic responses, and bulk electronic states in TCSCs with an order-two symmetry, including both symmorphic and nonsymmorphic symmetries in crystal. Under order-two symmetries, magnetic responses are classified into four types by one-dimensional (1D) topological invariants protecting Majorana Kramers pairs: (A) a TRS-protected $\mathbb{Z}_2$ invariant, (B) a symmorphic-symmetry-protected $\mathbb{Z}$ invariant, (C) a symmorphic-symmetry-protected $\mathbb{Z}_2$ invariant, and (D) a nonsymmorphic-symmetry-protected $\mathbb{Z}_2$ invariant, where types B and D depend on the number of Majorana Kramers pairs: Types B and D exhibit the same magnetic response for a single Majorana Kramers pair, whereas their magnetic responses are distinguishable for double ones. Such dependence on the number of Majorana Kramers pairs is crucial for TCSCs with nonsymmorphic symmetry. Furthermore, a $\mathbb{Z}_2$ invariant is related to Fermi-surface topology in the normal state, dubbed the Fermi surface criterion~\cite{Fu097001,Sato214526, Sato220504}. Extending it to those symmetry-protected $\mathbb{Z}_2$ invariants, we clarify that all types of magnetic responses ensured by $\mathbb{Z}_2$ invariants are determined only from the normal states and irreps of pair potentials, with the aid of additional symmetries. Thus, the anisotropic magnetic responses provide a clue to investigate the electronic states of TSCs/TCSCs, which provides a deep insight into the pairing mechanism of topological superconductivity and a way to detect and control Majorana fermions.

We also apply our generic theory to 
a model of UCoGe~\cite{Daido227001,Yoshida235105}, which has been recently proposed as a time-reversal-invariant TCSC at high pressure \cite{Hassinger073703, Slooten097003, Bastien125110, Manago020506,Cheung134516,Mineev104501}. The crystal of UCoGe has the space group $Pnma$ \cite{CANEPA1996225} so that our theory predicts a quadrupolar-shaped magnetic response induced by the glide plane. Carefully examining a surface state of the tight-binding model with a glide-plane symmetry, we demonstrate that double Majorana Kramers pairs on the $(0\bar{1}1)$ surface exhibit the quadrupolar-shaped magnetic response for the $B_{1u}$ pair potential, which offers a clue to identify Majorana Kramers pairs protected by the glide-plane symmetry.

The organization of this paper is as follows. 
First, in Sec.~\ref{Preliminary}, we introduce our notations of symmetry operations and discuss the AZ symmetry classes under order-two symmetries and associated 1D topological invariants.
In Sec.~\ref{Magnetic}, we classify possible magnetic responses of Majorana Kramers pairs protected by the 1D topological invariants and show a relation between the number of Majorana Kramers pairs and magnetic responses.
In Sec.~\ref{fermi}, generalizing the Fermi-surface criterion to the case of order-two symmetries, we derive several simplified formulae for symmetry-protected $\mathbb{Z}_2$ invariant and construct a connection between bulk electronic states and magnetic responses of Majorana Kramers pairs for $\mathbb{Z}_2$ invariants.
Finally, in Sec.~\ref{AppUCoGe}, we apply our theory to a superconducting state in UCoGe with the space group $Pnma$.
Summary and perspective are discussed in Sec.~\ref{conclusion}.

\section{Summary of the key results}
Before going to the main discussion, we briefly show summary of our results on the magnetic properties of Majorana Kramers pairs. 

\begin{description}
	\item[Four types of magnetic responses]\mbox{}\\
	Majorana Kramers pairs on TCSCs exhibit various magnetic responses, which are classified into four types, depending on their crystalline and superconducting symmetries, as summarized in Table \ref{fivetype}.
	An external uniform magnetic field can open a gap in Majorana Kramers pairs in an anisotropic form. 
	The number of Majorana Kramers pairs and whether the crystalline symmetry is symmorphic or nonsymmorphic are key ingredients for the magnetic responses. 
	\\
	For instance, a Majorana Kramers pair protected by a glide symmetry is gapped by an applied magnetic field perpendicular to the glide plane. This response is of type D in Table \ref{fivetype} with a single Majorana Kramers pair ($\#\mathrm{MKP}=1$). 
	
	\item[Fermi-surface criterion]\mbox{}\\
	The presence/absence of Majorana Kramers pairs is determined from the Fermi-surface topology in a given space group with a superconducting pair potential.
	Following the procedure shown in Figs.~\ref{Without time-reversal symmetry Fig}--\ref{glideZ2 Fig2}, we can systematically obtain the $\mathbb Z_2$ topological invariants corresponding to the parity of the number of Majorana Kramers pairs.
	Examples of this criterion are shown in Sec.~\ref{demo}.
\end{description}

\section{Preliminary}\label{Preliminary}

In this paper, we discuss Majorana Kramers pairs on a surface of 3D TCSCs. In the presence of TRS, Majorana Kramers pairs are, in general, protected by a 3D winding number and its parity.
The former is only defined for gapful superconductors, whereas the latter is defined even for gapless superconductors. We naively expect that adding a magnetic field makes both topological invariants ill-defined and destabilizes Majorana Kramers pairs. 
This scenario is true only for systems without additional symmetries.
Many candidate materials host crystalline symmetry keeping Majorana Kramers pairs even under a magnetic field in a particular direction. 
Our purpose is to provide a systematic procedure to identify such crystalline-symmetry-protected Majorana Kramers pairs and associated magnetic responses.
For this purpose, we focus on Majorana Kramers pairs at time-reversal-invariant momenta (TRIMs) that are classified by an order-two symmetry-protected 1D topological invariant~\cite{shiozaki14}. 
Note that the following discussions can be applied to lower-dimensional and nodal superconductors.

\subsection{Symmetry operations in superconducting states} 
First of all, we introduce symmetry operations that superconducting states host.
We start with the Bogoliubov-de Gennes (BdG) Hamiltonian in time-reversal-invariant superconductors:
\begin{align}
 H(\boldsymbol k) &= \mqty(
  h(\boldsymbol k) - \mu & \Delta(\boldsymbol k)
  \\
  \Delta(\boldsymbol k) & -h(\boldsymbol k) + \mu
 )
 \nonumber\\ &
 = \qty[h(\boldsymbol k)-\mu] \tau_z + \Delta(\boldsymbol k) \tau_x,
\end{align}
in the basis of $(c_{\boldsymbol k \uparrow}, c_{\boldsymbol k \downarrow}, c_{-\boldsymbol k \downarrow}^\dag, -c_{-\boldsymbol k \uparrow}^\dag)$, where $\uparrow$ and $\downarrow$ denote the up and down spins, respectively. 
The indices for the orbital and sublattice degrees of freedom are implicit. $h(\bm{k})$ is a normal Hamiltonian, $\Delta(\bm{k})$ a pair potential, and $\mu$ a chemical potential. $\tau_i$ $(i=x,y,z)$ are the Pauli matrices in the Nambu space.
The BdG Hamiltonian satisfies internal symmetries: time-reversal symmetry (TRS): $\Theta H(\vv{k})\Theta^{-1} = H(-\vv{k})$, particle-hole symmetry (PHS): $C H(\vv{k}) C^{-1} = -H(-\vv{k}), \ C = \tau_y \Theta$, and chiral symmetry given by the combination of TRS and PHS: $\{\Gamma,H(\vv{k})\}= 0, \ \Gamma = \Theta C = \tau_y$, where $\Theta$ is an antiunitary operator and satisfies $\Theta^2 =-1$.

When systems have a space-group symmetry ($\mathcal{G}$) in addition to the above internal symmetries, a normal Hamiltonian $h(\boldsymbol{k})$ satisfies 
\begin{align}
 D^{\dagger}_{\vv{k}}(g) h(\vv{k}) D_{\vv{k}}(g) = h(g\vv{k}), \ g \in \mathcal G, \label{eq:normalg}
\end{align}
with $D_{\boldsymbol{k}}(g)$ being a representation matrix of $g$, where a momentum $\boldsymbol{k}$ is transformed to $g\vv{k}$ under the action of $g$. 
We use the Seitz notation, i.e., $g = \{R_g | \boldsymbol{\tau}_g \}$ with $R_g$ rotation/reflection followed by translation $\boldsymbol{\tau}_g$. Here $\boldsymbol{\tau}_g$ is a primitive translation vector for symmorphic space groups and a non-primitive translation vector for nonsymmorphic space groups. 
Hereafter, the phase of representation matrices is fixed as $\Theta D_{\boldsymbol{k}}(g) \Theta^{-1} = D_{-\boldsymbol{k}}(g)$ in this paper. 

We focus our attention on order-two symmetry operations in spin--1/2 systems that satisfy $g^2 = \{{}^dE|R_g \boldsymbol{\tau}_g + \boldsymbol{\tau}_g\}$, where ${}^dE$ denotes the $2\pi$ rotation in the double group.  
Then, representation matrices at TRIM $\boldsymbol{k}_\Gamma$ satisfy
\begin{align}
 D^2_{\boldsymbol{k}_\Gamma}(g) = -e^{-i \boldsymbol{k}_\Gamma \cdot \qty(R_g \boldsymbol{\tau}_g + \boldsymbol{\tau}_g)} = \pm 1, 
 \label{eq:gpm}
\end{align}
which distinguishes nonsymmorphic space group operations from symmorphic ones.
$D^2_{\boldsymbol{k}_\Gamma}(g) = 1$ holds for $g$ being a glide plane or screw axis with a phase shift $\boldsymbol{k}_\Gamma \cdot 2 \boldsymbol{\tau}_g = \pi$. 
Otherwise, $D^2_{\boldsymbol{k}_\Gamma}(g) = -1$ always holds.

In superconducting states, space-group operations are defined in a similar way to those for a normal Hamiltonian, i.e., a pair potential is transformed as 
\begin{align}
 D^{\dagger}_{\vv{k}}(g) \Delta(\vv{k}) D_{\vv{k}}(g) = \chi (g) \Delta(g\vv{k}), \label{eq:deltag}
\end{align}
where $\chi(g)$ is the character of $g$ characterizing the irreps of pair potentials \cite{multi}. 
In particular, for order-two symmetries, $\chi(g)$ takes $\pm 1$. 
Using Eqs.~(\ref{eq:normalg}) and (\ref{eq:deltag}), the action of $g$ on the BdG Hamiltonian turns out to be
\begin{align}
 &\tilde{D}^{\dagger}_{\vv{k}}(g) H(\vv{k}) \tilde{D}_{\vv{k}}(g) = H(g\vv{k}), \label{eq:Dgaction}
 \\
 &\tilde{D}_{\vv{k}}(g) = \pmqty{D_{\vv{k}}(g) & 0 \\ 0 & \chi(g)D_{\vv{k}}(g)},
\end{align}
where the commutation relations between the symmetry operations $\tilde D_{\boldsymbol{k}}(g)$, $\Theta$, $C$, and $\Gamma$ satisfy
\begin{align}
 &\tilde D_{-\boldsymbol{k}}^\dag(g) \Theta \tilde D_{\boldsymbol{k}}(g) = \Theta,
 \label{TRS}  \\
&\tilde{D}^{\dagger}_{-\vv{k}}(g) C \tilde{D}_{\vv{k}}(g) = \chi(g)C,
\label{PHS} \\
&\tilde{D}^{\dagger}_{\vv{k}}(g) \Gamma \tilde{D}_{\vv{k}}(g) = \chi(g)\Gamma.
\label{chiral}
\end{align}
Therefore, Eqs.~(\ref{PHS}) and (\ref{chiral}) depend on $\chi(g)$, which manifests a connection between symmetry operators and irreps of pair potentials.

\subsection{Topological classification in 1D subspaces}

In the following, we review the topological classification in 1D systems with an order-two symmetry. Related topological classifications have been done in the previous studies~\cite{ueno13, chiu-yao-ryu, morimoto-furusaki, shiozaki14, Shiozaki195413, Benalcazar224503, Fang01944, Cornfeld075105, Shiozaki09354}. 

To begin with, we categorize a symmetry operation $g$ into two types: $U$, a symmetry operation that preserves the surface, and $P$, a symmetry operation that inverts the surface. The stability of Majorana Kramers pairs is ascribed to $U$, while $P$ indirectly affects them via bulk topological invariants. From Eq.~(\ref{eq:Dgaction}), the actions of $U$ and $P$ on the BdG Hamiltonians become
\begin{align}
 &\qty[\tilde{D}_{k_\perp}(U), H(k_\perp)] = 0, \label{eq:com_UH}
 \\
 &\tilde{D}_{k_\perp}^\dag(P) H(k_\perp) \tilde{D}_{k_\perp}(P) = H(-k_\perp), \label{eq:parity_sym}
\end{align} 
where $k_\perp$ and $\boldsymbol{k_\parallel}$ are momentum normal and parallel to the surface, respectively; $\boldsymbol{k_\parallel}$ is fixed at TRIMs.
Hereafter, we refer to $U$ and $P$ as a surface symmetry and a ``parity'' symmetry, respectively.  Note that ``parity'' stems from that $P$ behaves like inversion symmetry in a 1D subspace. In particular, we call a ``parity'' symmetry that satisfies $\tilde{D}_{k_\perp}^2(P)=1$ an ``inversion'' symmetry, because the original inversion operation satisfies this condition. 
We often find the ``inversion'' operation in systems with nonsymmorphic symmetry, such as glide and screw symmetries.   

Now we consider symmetry classes of the BdG Hamiltonians with a symmetry $U$, which stabilizes Majorana Kramers pairs. The effect of a ``parity'' symmetry will be discussed in Sec.~\ref{fermi}. Since the BdG Hamiltonians commute with $\tilde{D}_{k_\perp}(U)$, they can be decomposed in the eigenspaces of $\tilde{D}_{k_\perp}(U)$, i.e., $H \to H_+ \oplus H_-$, where the subscript $\pm$ denotes eigenvalues of $\tilde D_{k_\perp}(U)$. 
Under the decomposition, we can define AZ classes within the eigenspaces of $D_{k_\perp}(U)$, because a mixing term between $H_+$ and $H_-$ is forbidden by Eq.~(\ref{eq:com_UH}). 
$D_{k_\perp}(U)$ preserving the surface is independent of $k_\perp$, so we omit the subscript unless otherwise stated.

For an order-two symmetry, AZ classes depend only on the character $\chi(U)$ and the square $D^2(U)$: when $(\chi(U),D^2(U))=(1,1)$, it follows from Eqs.~(\ref{TRS}), (\ref{PHS}), and (\ref{chiral}) that $D(U)$ commute with TRS and PHS, and thus those symmetries keep within the same eigenspace. That is, $H_{\pm}$ belongs to class DIII. The corresponding 1D topological invariant is $\mathbb{Z}_2 \oplus \mathbb{Z}_2$, where two $\mathbb{Z}_2$ invariants are defined within each eigenspace. On the other hand, when $(\chi(U),D^2(U))=(1,-1)$, we obtain the same commutation relation among $D(U)$, TRS, and PHS, but $\Theta$ and $C$ are not preserved within the eigenspaces because $D^2(U)=-1$ has eigenvalues $\pm i$ and $\Theta$ and $C$ are antiunitary operators. 
Hence the eigenspaces have only the chiral symmetry, and then they belong to class AIII and are classified by a $\mathbb Z$ invariant. 
Similarly, we find class AII for $(\chi(U), D^2(U))=(-1,1)$ and class D for $(\chi(U), D^2(U))=(-1,-1)$, which are characterized by 1D topological invariants $0$ and $\mathbb{Z}_2$, respectively. 
Note that class D and AIII have a single topological invariant since $\Theta$ swaps $H_+$ for $H_-$ and their invariants are the same.
The resulting symmetry classes under symmetry $U$ are listed in Table \ref{symmetryclass}, where Majorana Kramers pairs appear when $(\chi(U),D^2(U))=(1,1)$, $(1,-1)$, and $(-1,-1)$. From Eqs.~(\ref{eq:gpm}) and (\ref{eq:deltag}), $D^2(U)$ takes $-1(1)$ for (non)symmorphic symmetry groups, and $\chi(U)$ is related to irreps of pair potentials. 
It is found that symmorphic symmetry groups lead to $\mathbb{Z}$ for a $U$-even pair potential and $\mathbb{Z}_2$ for a $U$-odd pair potential, whereas nonsymmorphic symmetry groups only to $\mathbb{Z}_2 \oplus \mathbb{Z}_2$ for a $U$-even pair potential.

\begin{table}
	\caption{AZ symmetry class and 1D topological invariant (Topo) of $H_{\pm}$ in the eigenspace of $\tilde{D}(U)$, where
		$\chi(U)$ and $D^2(U)$ are the character of $U$ and the square of representation matrix. 
	   $D^2(U)= -1 (1)$ indicates that $U$ is a (non)symmorphic symmetry group operation.}
	\begin{ruledtabular}
		\begin{tabular}{cccc}
			$\chi(U)$ & $D^2(U)$ & AZ & Topo
			\\
			\hline
			1 & 1 & DIII & $\mathbb Z_2 \oplus \mathbb Z_2$
			\\
			1 & $-1$ & AIII & $\mathbb Z$
			\\
			$-1$ & 1 & AII & 0
			\\
			$-1$ & $-1$ & D & $\mathbb Z_2$
			\\
		\end{tabular}
	\end{ruledtabular}
	\label{symmetryclass}
\end{table}

\subsection{1D topological invariants}

We here define the corresponding 1D topological invariants for each AZ class. We start with a $\mathbb{Z}_2$ topological invariant in the absence of symmetry $U$.  Since a single Majorana Kramers pair is solely protected by TRS, it is defined by
\begin{align}
 \nu _{\rm DIII} = \int \frac{dk}{2\pi} a_-(k) \mod 2, \label{eq:nuDIII} 
\end{align}
where $a_-(k)$ is the Berry connection in negative-energy states that satisfies $H(k) \ket{k-\alpha} =  - E_\alpha(k) \ket{k - \alpha}$:
\begin{align}
a_-(k)  = -i \sum_{\alpha}  \sum_{\zeta = \mathrm I, \mathrm{II}} \mel{k-\alpha \zeta}{\partial_k}{k - \alpha \zeta},
\end{align}
where $\alpha$ is the band index and $\zeta = \mathrm{I},  \mathrm{II}$ label a Kramers pair with the fixed gauge as
$ \Theta \ket{k - \alpha \mathrm I } = \ket{-k - \alpha \mathrm{II} }.$

When taking into account $U$, the stability of Majorana Kramers pairs can be ensured by symmetry-protected 1D topological invariants. 
First, we consider a topological invariant for $(\chi(U),D^2(U))=(1,-1)$, which belongs to class AIII. The corresponding 1D topological invariant is the magnetic winding number:
\begin{align}
 W[U] = \frac{i}{4\pi} \int dk \tr(\Gamma[U] H^{-1}\partial_k H) \in \mathbb Z, \label{eq:magnetic_winding}
\end{align}
where we use the magnetic chiral operator $\Gamma[U] = e^{i \phi} \Gamma \tilde D(U)$ instead of the usual chiral operator since the winding number in terms of $\Gamma$ becomes zero under TRS or PHS. The phase $\phi$ is chosen as $\Gamma[U]^2=1$. The bulk-edge correspondence manifests a one-to-one correspondence between the magnetic winding number and the number of the surface zero-energy states:
\begin{align}
W[U] = N_+ -N_- \label{eq:w-bbc}
\end{align}
where $N_\pm$ is the number of the surface zero energy states that are eigenstates of $\Gamma[U]$ with eigenvalues $\pm 1$. 
The physical implication of Eq.~(\ref{eq:w-bbc}) is that a pair of states with positive and negative eigenvalues of $\Gamma[U]$ is lifted by a small perturbation without breaking any symmetry due to $\{H,\Gamma[U]\}=0$. 
After all, only ($N_+ -N_-$) zero energy states remain stable.
In tight-binding BdG Hamiltonians with an open-boundary condition, Eq.~(\ref{eq:w-bbc}) is categorized only to three cases: $N_+ \neq 0 \land N_- =0$, $N_+ = 0 \land N_- \neq 0$, or $N_+ = N_- $~\cite{xiong17}.

In contrast, the cases of $(\chi(U),D^2(U))=(1,1)$ and $(-1,-1)$, respectively, belong to class DIII and D, whose 1D topological invariants are characterized by $\mathbb{Z}_2$. 
However, these invariants are different. 
To see this, we first consider the case of $(-1,-1)$ whose eigenspaces belong to class D. Thus, we can define a symmetry-protected $\mathbb{Z}_2$ invariant in class D:
\begin{align}
 \nu_{{\rm D}, \pm}[U] = \int\frac{dk}{\pi} a_{-}^\pm(k) \mod 2, \label{eq:nupmD}
 \end{align}
 with the Berry connection in the eigenspaces
 \begin{align}
 a_{-}^{\pm}(k) = -i \sum_{\alpha} \mel{k - \alpha \pm}{\pdv{k}}{k - \alpha \pm},
\end{align}
where $\ket{k-\alpha\pm}$ is an $\alpha$-th negative-energy state that satisfies
 $H_\pm(k) \ket{k-\alpha \pm} =  - E_\alpha^{\pm}(k) \ket{k - \alpha \pm}$ and $\pm$ represent eigenvalues of $U$. Due to TRS, the eigenspaces are related to each other, resulting in $\nu_{{\rm D},+}[U] = -\nu_{\rm{D},-}[U]$.

On the other hand,  for $(1,1)$, TRS keeps in the eigenspaces, i.e., those belong to class DIII, which hosts a symmetry-protected $\mathbb{Z}_2$ invariant in terms of a Kramers pair labeled by $\zeta = \mathrm {I}$ and $\mathrm{II}$. The corresponding $\mathbb{Z}_2$ invariant is defined in a similar way to Eq.~(\ref{eq:nuDIII}):
\begin{align}
 \nu_{{\rm DIII}, \pm}[U]  = \int \frac{dk}{2\pi} a_-^\pm(k) \mod 2, \label{eq:nupmDIII} 
\end{align}
with the Berry connection in the eigenspaces
\begin{align}
 a_-^\pm(k)  = -i \sum_{\alpha}  \sum_{\zeta = \mathrm I, \mathrm{II}} \mel{k-\alpha \zeta \pm}{\partial_k}{k - \alpha \zeta \pm},
\end{align}
where $\ket{k-\alpha\zeta\pm}$ is an $\alpha$-th negative-energy state that satisfies
 $H_\pm(k) \ket{k-\alpha \zeta \pm} =  - E_\alpha^{\pm}(k) \ket{k - \alpha \zeta \pm}$ and $\pm$ represent eigenvalues of $U$, and we fix the gauge as
$ \Theta \ket{k - \alpha \mathrm I \pm} = \ket{-k - \alpha \mathrm{II} \pm}.$
Note that $\nu_{{\rm DIII}, +}[U]$ and $\nu_{{\rm DIII}, -}[U] $ are independent of each other unless an additional constraint is imposed, so Majorana Kramers pairs are classified by $(\nu_{{\rm DIII}, +}[U], \nu_{{\rm DIII}, -}[U] ) \in \mathbb{Z}_2 \oplus \mathbb{Z}_2$.

\section{Magnetic response of Majorana Kramers pairs} 
\label{Magnetic}

We turn to magnetic responses of Majorana Kramers pairs, which are stabilized by one of those 1D topological invariants~(\ref{eq:nuDIII}), (\ref{eq:magnetic_winding}), (\ref{eq:nupmD}), and (\ref{eq:nupmDIII}). Since Majorana Kramers pairs are topological objects, we envision a magnetic field that makes the 1D topological invariants ill-defined is detrimental to their stability. 
This mechanism enables us to determine possible magnetic responses from the analysis of topological invariants; namely, a magnetic response of Majorana Kramers pairs occurs if a 1D topological number protecting them becomes unstable by adding a TRS breaking perturbation, e.g., the Zeeman terms, in the BdG Hamiltonian. 
Thus, the magnetic responses can be ascribed to Eqs.~(\ref{eq:nuDIII}), (\ref{eq:magnetic_winding}), (\ref{eq:nupmD}), and (\ref{eq:nupmDIII}); each case is categorized into types A, B, C, and D, respectively. 
In addition, double Majorana Kramers pairs appear when types B and D, characterized by $\mathbb{Z}$ and $\mathbb{Z}_2 \oplus \mathbb{Z}_2$, respectively. 
The number of Majorana Kramers pairs also affects magnetic responses, and the magnetic responses are different from those in a single Majorana Kramers pair.
The possible magnetic responses are summarized in Table \ref{fivetype}. 
A systematic classification including the $\mathbb{Z}_2$ invariants and double Majorana Kramers pairs has not been achieved in the previous works~\cite{shiozaki14, Dumitrescu245438, xiong17, kobayashi, yamazaki07939}. 
In the following, we show that magnetic responses are systematically determined from  those 1D topological invariants for a single Majorana Kramers pair. On the other hand, double Majorana Kramers pairs involve several couplings between Majorana Kramers pairs, which render magnetic responses complex. Thus, the magnetic responses depend not only on the 1D invariants but also on the details of material-dependent parameters in systems.    

\begin{table*}
 \caption{
 Four type of magnetic response of Majorana Kramers pairs, which are classified by relevant 1D topological invariants~(\ref{eq:nuDIII}), (\ref{eq:magnetic_winding}), (\ref{eq:nupmD}), and (\ref{eq:nupmDIII}). Second, third, fourth and fifth columns represent surface symmetries that protect Majorana Kramers pairs, the character of $U$, the square of $D(U)$, the number of Majorana Kramers pairs, and the energy gap of Majorana Kramers pairs under an applied magnetic field, respectively. For types A and C, a Majorana Kramers pair is protected the $\mathbb{Z}_2$ invariant, so only a single Majorana Kramers pair appears, whereas for type B and D, Majorana Kramers pairs are protected by $\mathbb{Z}$ and $\mathbb{Z}_2 \oplus \mathbb{Z}_2$, respectively. Interestingly, types B and D become distinguishable for double Majorana Kramers pairs.
 }
 \begin{ruledtabular}
  \begin{tabular}{clcccl}
  Type & Surface symmetry & $\chi(U)$ & $D^2(U)$ & \#MKP & $E_{\mathrm{M}}(\boldsymbol{B})$
  \\
  \hline
  A & TRS only                            & $-$ & $-$  & $1$  & Eq.~(\ref{eq:general_form})
 \\
  B & Magnetic chiral $\Gamma[U]$ & $1$ & $-1$  & $1$  & Eq.~(\ref{eq:general_form}) with $f (U\bm{B}) = f (\bm{B})$
  \\
     &                                        & $1$ & $-1$  & $N \ge 2$  & Eq.~(\ref{eq:diagonal_form}) with $f_i (U\bm{B}) = f_i (\bm{B})$ and $g_i (U\bm{B}) = -g_i (\bm{B})$
  \\
  C & Symmorphic $U$                 & $-1$ & $-1$ & $1$ & Eq.~(\ref{eq:general_form}) with $f (U\bm{B}) = -f (\bm{B})$
  \\
  D & Nonsymmorphic $\Theta[U]$ & $1$  & $1$   & $1$  & Eq.~(\ref{eq:general_form}) with $f (U\bm{B}) = f (\bm{B})$
  \\
     &                                        & $1$ & $1$   & $2$  & Eq.~(\ref{eq:diagonal_form}) 
 \end{tabular}
 \end{ruledtabular}
\label{fivetype}
\end{table*}

\subsection{A single Majorana Kramers pair}
\label{sec:single}
First of all, we discuss a general form of magnetic responses for a single Majorana Kramers pair. We start with a surface effective theory describing couplings between Majorana fermions. Let us define $N$ Majorana Kramers pairs $\gamma_1, \cdots \gamma_{2N}$ satisfying $\gamma_i^{\dagger} = \gamma_i$ and $\{\gamma_i,\gamma_j\}=\delta_{ij}$. 
The time-reversal operation changes $\gamma_{2n-1}$ to $\gamma_{2n}$ and $\gamma_{2n}$ to $-\gamma_{2n-1}$ since $\Theta^2=-1$. 
Then, the coupling between $N$ Majorana Kramers pairs is given by 
\begin{align}
J = \frac{1}{2}\bm\gamma^{\mathrm T} A \bm\gamma,
 \
 A^{\mathrm T} = - A, \label{eq:efftheory}
\end{align}
where $\bm{\gamma}= (\gamma_1,\dots,\gamma_{2N})^{\mathrm T} $ and $J^{\dagger} = J$. Equation~(\ref{eq:efftheory}) can be applicable in the low-energy regime, i.e., the energy scale of TRS-breaking terms is much smaller than the superconducting gap.
 
Provided that a single Majorana Kramers pair exists at a TRIM in a surface BZ, one is described by $\bm{\gamma}= (\gamma_1,\gamma_{2})^{\mathrm T} $. 
$A$ is the antisymmetric matrix, uniquely represented by $A = s_y$, where $s_i \ (i=0,x,y,z)$ denotes the $2 \times 2$ identity matrix and the Pauli matrices. 
In addition, TRS imposes that Eq.~(\ref{eq:efftheory}) is invariant under $\gamma_1 \rightarrow \gamma_2$ and $\gamma_2 \rightarrow -\gamma_1$. One finds that $J$ is a time-reversal-odd (magnetic) operator and coupled to a magnetic field $\boldsymbol{B}$.
Thus, a surface Hamiltonian in terms of a single Majorana Kramers pair under a magnetic field $\boldsymbol{B}$ is of form $H_{\mathrm{MF}} = A f(\boldsymbol{B})$ and its energy spectrum $E_{\text{M}}(\boldsymbol{B}) = \pm f(\boldsymbol{B})$, where $f(\boldsymbol{B})$ is an analytic odd function of $\bm{B}$, explicitly shown as~\cite{Dumitrescu245438}
\begin{align}
 f(\vv{B}) = \sum_{i} \rho_i B_i + \sum_{i,j,k}\rho_{ijk} B_iB_jB_k + \order{B^5},
\label{eq:general_form}
\end{align}
where the coefficients $\rho_i$ and $\rho_{ijk}$ depend on the details of the system. 
We emphasize that $A$ is fixed due to the property of Majorana operators and TRS, and only a magnetic response is realized. 
As a corollary of the fact, we find a one-to-one correspondence between anisotropic magnetic responses and bulk physical quantities, as we shall see in Sec.~\ref{fermi}.

\subsubsection{Type A}
Now we see the connection between the 1D topological invariants and Eq.~(\ref{eq:general_form}), and readily find that a magnetic response for Majorana Kramers pairs protected by Eq.~(\ref{eq:nuDIII}), i.e., type A, is of the form~(\ref{eq:general_form}) since there is no symmetry constraint. On the other hand, Eqs.~(\ref{eq:magnetic_winding}), (\ref{eq:nupmD}), and (\ref{eq:nupmDIII}) provide an additional constraint of $U$ on Eq.~(\ref{eq:general_form}), as explained hereafter.

\subsubsection{Type B}
Firstly, we consider a Majorana Kramers pair protected by the magnetic winding number when surface symmetry $U$ obeys $(\chi(U), D^2(U))=(1,-1)$.
In this case, Eq.~(\ref{eq:magnetic_winding}) is well-defined as long as the BdG Hamiltonian satisfies magnetic chiral symmetry $\{H,\Gamma[U]\}=0$. 
Hence, if a TRS-breaking term, such as the Zeeman and vector potential terms, commutes with $\Gamma[U]$, the magnetic chiral operator is no longer preserved and Eq.~(\ref{eq:magnetic_winding}) becomes ill-defined. 

In addition, Majorana Kramers pairs become unstable under applied magnetic fields in a specific direction. 
To see this concretely, we consider the Zeeman magnetic term $H_{\rm mag} \propto \bm{B} \cdot \bm{s}$, where $s_i$'s are the Pauli matrices in the spin space. 
Since $[H_{\rm mag }, \Gamma ]=0$, the condition for the instability is $[H_{\rm mag }, \tilde{D}(U) ]=0$. 
For instance, let $\tilde D(U) = i \bm{n} \cdot \bm{s} \tau_0$ be a twofold rotation operation in the direction $\bm{n}$. 
Then, the Zeeman term in the $\bm{n}$ direction $H_{\mathrm{mag}} \propto B \boldsymbol{n} \cdot \boldsymbol{s} \tau_0$ leads to the instability condition $[H_{\rm mag}, \tilde D(U)]=0$. 
The same commutation relation is found when $U$ is a reflection or glide operation whose symmetric plane is perpendicular to $\bm{n}$. 
In both cases, the energy gap induced by a magnetic field $\bm{B}$ can be proportional to $\bm{n} \cdot \bm{B}$, which response has been known as the Ising anisotropy~\cite{shiozaki14, xiong17}. 
Such an anisotropy would be observed in superfluid $^3$He-B phase~\cite{Leggett331,Chung235301,Nagato123603,Mizushima12,Mizushima022001}, doped topological insulator~\cite{Fu097001,kobayashi}, and the TRS-preserving $E_{1u}$ state of UPt$_3$~\cite{Sauls113,Tsutsumi074717,Tsutsumi113707,Mizushima184506,UPt3}. 
The same argument holds true for an arbitrary TRS--breaking perturbation. 
%
Trigonal and hexagonal crystals potentially show an exotic magnetic response of Majorana Kramers pairs, an octupole-shaped energy gap $E_{\mathrm{M}}(\boldsymbol{B}) \propto B_x^3 - 3 B_x B_y^2$ as a leading contribution when reflection symmetry protects a Majorana Kramers pair~\cite{kobayashi}. 

In general, Eq.~(\ref{eq:general_form}) for type B satisfies    
\begin{align}
	f (U\bm{B}) = f(\bm{B}), \label{eq:mag_typB}
\end{align}
where $U\bm{B}$ represents a rotation of axial vectors under $U$, e.g., $(B_x,B_y,B_z) \to (-B_x,-B_y,B_z)$ when $U$ is the twofold rotation in the $z$ direction.  

\subsubsection{Type C}
Secondly, we discuss magnetic responses of a Majorana Kramers pair protected by Eq.~(\ref{eq:nupmD}), which is a $\mathbb{Z}_2$-invariant-induced magnetic response.
Under the surface symmetry $U$, the BdG Hamiltonian is split into eigenspaces of $U$, $H_+ \oplus H_-$, and each eigenspace belongs to class D since $(\chi(U), D^2(U))=(-1,-1)$. 
Hence, the stability of a Majorana Kramers pair relies only on $U$ and $C$. 
Since PHS ($C$) remains intact under a magnetic field, the instability comes from the breaking of $U$, i.e., a TRS-breaking term that anticommutes with $U$ kills a Majorana Kramers pair. 
For instance, consider a twofold rotation, reflection, or glide operation $U = i \bm{n} \cdot \bm{s}$ and apply the Zeeman magnetic field $H_{\mathrm{mag}} \propto \boldsymbol{B} \cdot \boldsymbol{s} \tau_0$ to the Hamiltonian. 
Then, the anticommutation relation $\{H_{\mathrm{mag}}, U\}=0$ is satisfied for $\bm{B} \perp \bm{n}$. 
Thus, the magnetic response stands in stark contrast to type B, and generally, Eq.~(\ref{eq:general_form}) for type C satisfies    
\begin{align}
	f (U\bm{B}) = -f (\bm{B}). \label{eq:mag_typC}
\end{align}

\subsubsection{Type D}
Thirdly, a different $\mathbb{Z}_2$-invariant-induced magnetic response appears for a Majorana Kramers pair protected by Eq. (\ref{eq:nupmDIII}). 
Similarly to the above case, the BdG Hamiltonian is transformed into a form of $H_+ \oplus H_-$ with respect to eigenspaces of $U$, but $H_{\pm}$ are of class DIII since $(\chi(U), D^2(U))=(1,1)$. 
Such a situation happens when $U$ is a glide operator and a TRIM $\bm{k}_{\Gamma}$ satisfies $\bm{k}_{\Gamma} \cdot 2\bm{\tau}_U =\pi$. 
TRS breaks owing to a magnetic field, but still, $U$ supports the stability of the Majorana Kramers pair. 
Since a magnetic field perpendicular to the glide plane preserves $U$, the AZ class in the eigenspaces changes from class DIII to D, and the $\mathbb{Z}_2$ invariant (\ref{eq:nupmDIII}) is trivialized, leading to the disappearance of the Majorana Kramers pair. 
On the other hand, when an applied magnetic field is parallel to the glide plane, a TRS--breaking term anticommutes with $U$. 
Then, the magnetic--glide symmetry, defined by the combination of the glide plane and time reversal, $\Theta[U] \equiv \Theta \tilde{D} (U)$, is preserved, and plays a role of an emergent TRS under an applied magnetic field. 
Therefore, the Majorana Kramers pair remains stable under a magnetic field in this direction, and the magnetic response of type D is of the form~(\ref{eq:mag_typB}) with $\bm{n}$ being perpendicular to the glide plane. 
That is, the magnetic response is similar to that of type B. 
Interestingly, types B and D are distinguishable when double Majorana Kramers pairs are considered, as we shall discuss in the following.

\subsection{Double Majorana Kramers pairs}
\label{sec:double}
So far, we have seen the magnetic responses of a single Majorana Kramers pair described by Eq.~(\ref{eq:general_form}). 
We now shift our focus to multiple Majorana Kramers pairs, which will appear in the case of $(\chi(U),D^2(U))=(1,1)$ and $(1,-1)$. 
We envision that the magnetic responses become more complicated than those for a single Majorana Kramers pair. 
Yet, it is possible to determine a general form of magnetic responses and symmetry constraint on it. 
For simplicity, we focus on double Majorana Kramers pairs as a minimal setup. 

To see a general form of magnetic responses for double Majorana Kramers pairs, we recall the effective theory Eq.~(\ref{eq:efftheory}). 
Double Majorana Kramers pairs are given by $\bm{\gamma}= (\gamma_1,\gamma_{2}, \gamma_{
3}, \gamma_{4})^{\mathrm T} $, where $\gamma_{2n-1}$ and $\gamma_{2n}$ form a Kramers pair. 
Bearing the fact that $A$ is an antisymmetric matrix in mind, possible coupling terms between Majorana fermions are represented by six matrices: $A_1 = s_y \tau_0$, 
$A_2 = s_y \tau_z$, $A_3 = s_0 \tau_y$,  $A_4 = s_y \tau_x$,  $B_1 = s_z \tau_y$, and $B_2 = s_x \tau_y$, where $\tau_i$ ($i=0,x,y,z$) are the identity matrix and the Pauli matrices describing the coupling between different Majorana Kramers pairs. 
Time reversals of $A_i$ and $B_i$ are given by $-A_i$ and $B_i$, so we call them magnetic and electric coupling terms, respectively. 
Using these matrices, a surface Hamiltonian that describes the coupling between double Majorana Kramers pairs and magnetic fields can be constructed as
\begin{align}
 H_{\text{MF}} = \sum_{i=1}^{4} A_i f_i(\vv B) + \sum_{i=1}^{2} B_i g_i(\vv{B}),
 \label{HMF_double}
\end{align}
where $f_i(\bm{B})$ [$g_i (\bm{B})$] is an analytic odd (even) function of $\bm{B}$.
By diagonalizing $H_{\mathrm{MF}}$, one gets the energy spectrum of double Majorana Kramers pairs as
\begin{align}
E_{\text{M}}(\boldsymbol{B}) = \pm \left(\sqrt{f_1^2+f_3^2+f_4^2} \pm \sqrt{f_2^2 + g_1^2 + g_2^2} \right). \label{eq:diagonal_form}
\end{align}
Since mainly the low--energy gap can be contributed to the magnetic response, we choose the low--energy branch of Eq.~(\ref{eq:diagonal_form}) and approximate it in the lowest order of $\bm{B}$. 
Then, a general form of double Majorana Kramers pairs is of form 
\begin{align}
 E_{\text{M}}(\boldsymbol{B}) \sim \pm \left(\sqrt{\sum_{ij} \rho_{ij} B_i B_j} - \sqrt{\sum_{ij} \rho'_{ij} B_i B_j} \right),
 \label{eq:general_form_double}
\end{align}
where $\rho_{ij}$ and $\rho_{ij}^{\prime}$ are parameters, determined from the details of the systems.
In the following, we discuss the influence of the surface symmetry $U$ on Eq.~(\ref{eq:general_form_double}). 

\subsubsection{Type D}
For $(\chi(U),D^2(U))=(1,1)$, double Majorana Kramers pairs appear when two $\mathbb{Z}_2$ invariants, i.e., $(\nu_+[U], \nu_-[U] ) \in \mathbb{Z}_2 \oplus \mathbb{Z}_2$ in Eq.~(\ref{eq:nupmDIII}), are both nontrivial. 
As discussed above, we have the magnetic glide symmetry $\Theta[U]$, but it does not guarantee the stability of double Majorana Kramers pairs. 
Thus, there is no symmetry constraint on Eq.~(\ref{eq:general_form_double}), and the magnetic response of type D is Eq.~(\ref{eq:general_form_double}) itself.
A typical example is quadrupole-shaped anisotropy which will be shown in Sec.~\ref{AppUCoGe}.

\subsubsection{Type B}
On the other hand, for $(\chi(U),D^2(U))=(1,-1)$, i.e., type B, the relevant topological invariant is Eq.~(\ref{eq:magnetic_winding}), i.e., $W[U] \in \mathbb{Z}$, which protects $N$ ($=|W[U]|$) Majorana Kramers pairs. 
Accordingly, $N$ Majorana Kramers pairs are stable unless a TRS--breaking term breaks the magnetic chiral symmetry. 
In the following, we pay attention to double Majorana Kramers pairs. 

To break the magnetic chiral symmetry, a symmetry constraint on the effective Hamiltonian is imposed as
\begin{align}
 [\Gamma[U], A_i] = 0, 
 \
 [\Gamma[U], B_i] = 0,
\end{align}
leading to $[\Gamma[U], H_{\mathrm{MF}} ] = 0$.
$A_i$ and $B_i$ are magnetic and electric terms hence they must satisfy
\begin{align}
 \qty[\Gamma, A_i] = 0,
 \
 \qty{\Gamma, B_i} = 0.
\end{align}
Combining the above two relations, we obtain
\begin{align}
 \qty[\tilde D(U), A_i] = 0,
 \
 \qty{\tilde D(U), B_i} = 0.
\end{align}
Therefore, the coefficients in the effective Hamiltonian are constrained to be
\begin{align}
 &f_{i} (U\bm{B}) = f_{i} (\bm{B}), \label{eq:mag_typeBd1}\\
 &g_{i} (U\bm{B}) = -g_{i} (\bm{B}), \label{eq:mag_typeBd2}
\end{align} 
due to $U$ symmetry, $\tilde {D}^\dag(U) H_{\boldsymbol{B}} \tilde {D}(U) = H_{U \boldsymbol{B}}$.
The condition (\ref{eq:mag_typeBd1}) is similar to Eq.~(\ref{eq:mag_typB}), while (\ref{eq:mag_typeBd2}) takes a minus sign since $B_i$ are electric operators. 
To see how Eqs.~(\ref{eq:mag_typeBd1}) and (\ref{eq:mag_typeBd2}) affect Eq.~(\ref{eq:general_form_double}), we assume that $U$ is a twofold rotation symmetry in the $z$ direction. 
Then, $f_i$ and $g_i$ take the form:
\begin{align}
 &f_i (\bm{B}) = \rho_{i,z} B_z + \mathcal{O}(B^3), \label{eq:mag_typeBdeg1} \\
 &g_i (\bm{B}) = \rho_{i,xz} B_x B_z +\rho_{i,yz} B_y B_z + \mathcal{O}(B^4). \label{eq:mag_typeBdeg2}
\end{align}
 Substituting Eqs.~(\ref{eq:mag_typeBdeg1}) and (\ref{eq:mag_typeBdeg2}) into Eq.~(\ref{eq:diagonal_form}), it follows that
 \begin{align}
  E_{\text{M}}(\boldsymbol{B}) &\sim \pm B_z \left( \sqrt{\alpha} - \sqrt{\beta} \right) \label{eq:mag_typeBdc2z}
 \end{align}
where $\alpha = \rho_{1,z}^2+\rho_{3,z}^2+\rho_{4,z}^2$ and $\beta = \rho_{2,z}^2 + \sum_{i=1,2}(\rho_{i,xz}B_x +\rho_{i,yz}B_y)^2$. 
Thus, the energy gap is proportional to $B_z$ and exhibits the Ising anisotropy like a single Majorana Kramers pair when $|\bm{B}| \ll 1$.
We emphasize that Eqs.~(\ref{eq:general_form}) and (\ref{eq:general_form_double}) are qualitatively different in that the latter includes an electric response and can potentially react to electric perturbations.

\subsection{Comparison with a single Dirac fermion}

It is interesting to compare the above results with those of a single Dirac fermion on the surface of a topological insulator, which has three components $s_x$, $s_y$, and $s_z$ of spin due to the absence of particle-hole symmetry. 
The spin operators are defined by $s_i = \sum_{ss'} c^\dag_s (s_i)_{ss'} c_{s'}$ in terms of complex-fermion operators $c_s$ and $c^\dag_s$, not Majorana ones $\gamma_i$. 

\subsubsection{Gap opening}
Two-fold degeneracy on the Dirac point is lifted by the Zeeman effect, $H_{\mathrm{Dirac}} = \sum_i \rho_i B_i s_i$, as 
\begin{align}
	E_{\mathrm{Dirac}}(\boldsymbol{B}) = \sqrt{\rho_x^2 B_x^2 + \rho_y^2 B_y^2 + \rho_z^2 B_z^2},
\end{align}
with nonzero coefficient $\rho_x$, $\rho_y$, and $\rho_z$. 
$E_{\mathrm{Dirac}}(\boldsymbol{B})$ takes a nonzero value for an arbitrary magnetic field $\boldsymbol{B}$ and is distinct from that of a Majorana Kramers pair, $E_{\mathrm{Dirac}}(\boldsymbol{B}) \ne E_{\mathrm M}(\boldsymbol{B}) \sim \boldsymbol{\rho} \cdot \boldsymbol{B}$ of Eq.~(\ref{eq:general_form}). 

\subsubsection{Shift of Dirac point}
When the surface and the applied magnetic field respect mirror-reflection symmetry, the Dirac point shifts to a momentum $\boldsymbol{k}_0$ on the reflection-invariant plane \cite{fu09}, $E_{\mathrm{Dirac}}(\boldsymbol{B}, \boldsymbol{k}_0)=0$.
The shifted Dirac point is protected by the mirror Chern number \cite{teo10}.
To make the discussion more concrete, the mirror is set on the $xz$ plane. 
The Dirac point shifts on the $xz$ plane, i.e., $\boldsymbol{k}_0 = (a, 0)$, for $\boldsymbol{B} = (0, B_y)$. 
Such a shift is prohibited for a Majorana Kramers pair: The zero-energy state of a Majorana Kramers pair is always pinned on the TRIM because of particle-hole symmetry, $E_{\mathrm{M}}(\boldsymbol{B},\boldsymbol{k}) = -E_{\mathrm{M}}(\boldsymbol{B}, -\boldsymbol{k})$. 
On the contrary, for double Majorana Kramers pairs, the zero-energy point can be moved while maintaining particle-hole symmetry. 

\subsection{Relation to quantum tunnelings}

In the present section, we have mentioned that a chiral Majorana fermion has no physical operator and a single Majorana Kramers pair has no electric operator. Additionally, we have implied that double Majorana Kramers pairs have two operators coupled to time-reversal-invariant (electric) degrees of freedom as $\gamma_i \gamma_j \rho$, where $\rho$ denotes a classical or quantum electric quantity. 

Besides, these Majorana fermions $\gamma_i$ can be coupled to complex fermions $c_j$ via the quantum tunneling as $\gamma_i (c_j + c_j^\dag)$, which leads to tunneling conductance between the TCSC and a normal metal \cite{tanaka95, law09}. 
The Josephson coupling between Majorana fermions is also possible since it is a quantum effect arising from a nonuniform phase $\phi(\boldsymbol{r})$ of pair potential $\Delta(\boldsymbol{r}) = |\Delta| e^{i \phi(\boldsymbol{r})}$. 
This results in an anomalous Josephson effect \cite{wang11, chung13}. 

\section{Some relations among bulk electronic structures, 1D \texorpdfstring{$\mathbb{Z}_2$}{Z2} topological invariants, and surface magnetic responses}
\label{fermi}

Anisotropic magnetic response of Majorana Kramers pairs is a direct consequence of the bulk-boundary correspondence between surface zero modes and bulk topological invariants. 
This fact tells us that we can determine surface magnetic responses only from bulk electronic structures. 
To this end, we discuss a relation between surface magnetic responses and the bulk electronic structures through the symmetry-protected 1D topological invariants. 
Such a relation has been discussed in the case of the $\mathbb{Z}$ invariant, i.e., type B \cite{xiong17, kobayashi}, where the magnetic chiral symmetry plays a key role. 
Here we extend it to the $\mathbb Z_2$ topological invariants in types A, C, and D.
Especially, those $\mathbb Z_2$ invariants can be reduced to a simple formula with the help of a ``parity'' symmetry, an ``inversion'' symmetry, and an additional surface symmetry. 

We briefly summarize the results as follows.
We start with a relation between the $\mathbb Z_2$ topological invariant and the number of Fermi surfaces, which is called the Fermi-surface criterion, for TRS-breaking systems in Sec.~\ref{woTRS}.
This criterion tells us that an odd number of Fermi surfaces lead to topologically nontrivial superconductivity except for even-``parity'' pairings. 
The Fermi-surface criterion is extended to time-reversal-symmetric systems with crystalline symmetry, i.e., the $\mathbb Z_2$ invariants in types A, C, and D, in Sec.~\ref{wTRS} thanks to ``inversion'' symmetry including glide plane and screw axis.  
We also illustrate flowcharts of the Fermi surface criterion for each case, which will offer a guideline for searching for the topological invariants and a connection between surface magnetic responses and bulk electronic structures.

\subsection{Without TRS}
\label{woTRS}

\subsubsection{Without symmetry}
\label{Without time-reversal symmetry}
First of all, we revisit the Fermi-surface criterion for systems without symmetry (class D). 
We assume the weak-coupling condition that $H(k=0, \pi)$ is deformed continuously to $H(k=0, \pi)|_{\Delta=0}$. 
Under the assumption, the $\mathbb Z_2$ invariant is obtained to be
\begin{align}
 \nu_{\text{D}} = \# \mathrm{FS} \mod 2, \label{eq:FScriterion}
\end{align}
where $\# \mathrm{FS}$ indicates the number of the Fermi surfaces between $k=0$ and $\pi$. 
This Fermi-surface criterion has been well-known, but the proof needs spatial-inversion symmetry. 
In the following, we provide an alternative proof of it, without using spatial-inversion symmetry. 

We fix the gauge as $C \ket{k \pm \alpha} = \ket{-k \mp \alpha}$.
In this choice, the particle-hole symmetry of $a_\pm(k)$ is represented by $a_\pm(k) = a_\mp^*(-k)$.
Then, the 1D topological invariant in class D is rewritten as
\begin{align}
 \nu_{\text{D}} &= -i \int_0^\pi \frac{dk}{\pi} \tr u^\dag(k) \pdv{u(k)}{k}
 = -\frac{i}{\pi} \ln \frac{\det u(\pi)}{\det u(0)} \hspace{-1.5mm} \mod 2,
\end{align}
where we used $\tr \ln u(k) = \ln \det u(k)$, and the unitary matrix $u(k) \equiv (\ket{k +1} \ \cdots \ \ket{k + N} \ C \ket{-k+1} \ \cdots \ C\ket{-k +N})$ with $\ket{k + n}$ the $n$-th eigenvector for a positive eigenvalue of BdG Hamiltonian. 
Now we impose the weak-coupling assumption.  
The eigenvector for $\Delta(k_\Gamma) \to 0$ on a TRIM $k_{\Gamma}$ is given by
\begin{align}
\ket{k_\Gamma + \alpha} = 
\begin{cases}
\pmqty{
	v_\alpha(k_\Gamma)
	\\
	0
}
& \qfor \epsilon_\alpha(k_\Gamma) > \mu,
\\[1em]
\pmqty{
	0
	\\
	i s_y v^*_\alpha(k_\Gamma)
}
& \qfor \epsilon_\alpha(k_\Gamma) < \mu,
\end{cases}
\end{align}
where $\epsilon_\alpha(k)$ and $v_\alpha(k)$ are the normal-state energy and eigenvector, which is the solution of 
$h(k) v_\alpha(k) = \epsilon_\alpha(k) v_\alpha(k)$. 
Here the particle-hole conjugate is expressed by $C = \tau_y \Theta = \tau_y s_y \mathcal K$. 
For $\mu \to -\infty$,  we have
\begin{align}
 u(k_\Gamma) 
 &= \pmqty{
   v(k_\Gamma) & 0
   \\
   0 & is_y v^*(k_\Gamma)
 },
 \
 v(k) = (v_1(k) \ \cdots \ v_N(k)).
\end{align}
Then the determinant is given by $\det u(k_\Gamma) = |\det v(k_\Gamma)|^2 = 1$.
When the chemical potential is located at $\epsilon_1(k) < \mu < \epsilon_2(k)$, the eigenvectors are given by
\begin{align}
 u(k_\Gamma) = \pmqty{
 	0 & v'(k_\Gamma)  & v_1(k_\Gamma) & 0 
 	\\
  is_y v_1^*(k_\Gamma) & 0  & 0 & is_y v'^*(k_\Gamma) 
 },
\end{align}
where $v'(k) = (v_2(k) \ \cdots \ v_N(k))$.
By exchanging the 1st and $(N+1)$-th columns, one finds
\begin{align}
 \det u(k_\Gamma)
 = - \det\pmqty{v(k_\Gamma) & 0 \\ 0 & is_yv^*(k_\Gamma)} = -1.
\end{align}
Repeating this procedure, we obtain the determinant for arbitrary $\mu$ as $\det u(k_\Gamma) = (-1)^{N_{\mathrm F}(k_\Gamma)}$,
where $N_{\mathrm F}(k)$ is the number of states below the Fermi energy at momentum $k$. 
As a result, the $\mathbb Z_2$ invariant is given by the Fermi-surface criterion:
\begin{align}
 \nu_{\text{D}} = N_{\mathrm{F}}(\pi) - N_{\mathrm{F}}(0) = \#\mathrm{FS} \mod 2. 
\end{align}

\subsubsection{With ``parity" symmetry}
\label{parity}
\begin{figure}
	\centering
	\includegraphics[scale=0.27]{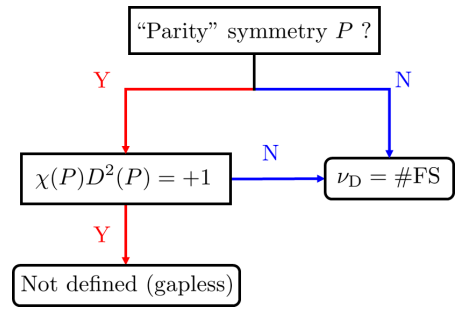}
	\caption{Fermi-surface criterion for systems without TRS, where
	``Parity" symmetry is defined by $D^{\dagger}(P)h(-k_{\perp},\bm k_{\parallel})D(P)=h(k_{\perp},\bm k_{\parallel})$ with $D^2(P)=\pm1$.}
	\label{Without time-reversal symmetry Fig}
\end{figure}

In addition,  ``parity'' symmetry [Eq.~(\ref{eq:parity_sym})]  is influential to the Fermi-surface criterion.
When a ``parity'' symmetry exists, the BdG Hamiltonian satisfies $\{C\tilde{D}(P), H(k)\}=0$ in conjunction with PHS. That is, the symmetry class of $H(k)$ in the 0D system changes to class D for $\chi(P) D^2(P) = 1$ and class C for $ \chi(P) D^2(P) = -1$, where we use Eq. (\ref{PHS}).  
As a result, a non-degenerate Fermi surface cannot be gapped when  $\chi(P)D^2(P)=1$, i.e., the even--``parity'' pair potential, 
because it is protected by the fermion parity for arbitrary $k$ \cite{agterberg17, kobayashi14, zhao16}. 
Note that when the pair potential is sufficiently large to hybridize states on and away from the Fermi energy, a topologically trivial gapped phase ($\nu_{\rm D} =0$) emerges. 
On the other hand, there is no constraint in the case of $\chi(P)D^2(P)=-1$. Hence, Eq.~(\ref{eq:FScriterion}) is always valid.
The Fermi-surface criterion for systems without TRS is summarized in Fig. \ref{Without time-reversal symmetry Fig}.

\subsection{With TRS}
\label{wTRS}

\subsubsection{Type A}
\label{With time-reversal and inversion symmetries}
\begin{figure}
	\centering
	\includegraphics[scale=0.27]{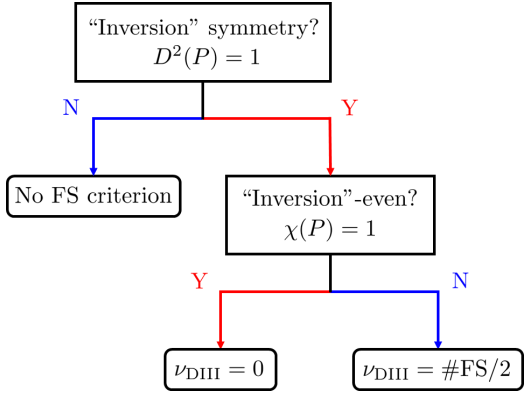}
	\caption{Fermi-surface criterion for type A, where  
	 ``Inversion" symmetry is defined by $D^{\dagger}(P)h(-k_{\perp},\bm k_{\parallel})D(P)=h(k_{\perp},\bm k_{\parallel})$ with $D^2(P)=1$.
	}
	\label{With time-reversal and inversion symmetries Fig}
\end{figure}

The Fermi-surface criterion can be applied to time-reversal-invariant systems when spatial-inversion symmetry exists \cite{Sato094504,Sato214526,Zhang134508,Sato220504,Fu097001}. 
Here we extend the Fermi-surface criterion to systems with ``inversion'' symmetry, where ``inversion'' ($P$) satisfies Eq.~(\ref{eq:parity_sym}) and $D^2(P)=1$. Note that $P$ refers not only to spatial-inversion but also to glide-plane or screw-axis symmetry, which satisfies the same condition in a 1D subspace.
In the presence of ``inversion'' symmetry, energy bands are twofold degenerate due to the Kramers theorem, and the topology of TSCs is related to an eigenvalue of ``inversion'' symmetry at TRIMs. Following the procedures~\cite{Sato220504,Fu097001}, Eq.~(\ref{eq:nuDIII}) can be simplified to
\begin{align}
 \nu_{\text{DIII}} = 
 \begin{cases}
  0 & \qfor \chi(P)=1,
  \\
  \displaystyle\frac{N_{\mathrm{F}}(\pi)-N_{\mathrm{F}}(0)}{2} = \frac{\#\mathrm{FS}}{2} & \qfor \chi(P)=-1,
  \label{nuDIII}
 \end{cases}
\end{align}
with modulo 2. Thus, the surface magnetic response of type A is intrinsically related to the number of bulk Fermi surfaces.  
Note that a ``parity" symmetry with $D^2(P)=-1$ does not lead to the Fermi-surface criterion since the $P$ exchanges Kramers partners. 
The Fermi-surface criterion for type A and the relevant flowchart are summarized in Fig. \ref{With time-reversal and inversion symmetries Fig}.

\subsubsection{Type C}
\label{TRSsymmorphic}
\begin{figure*}
	\centering
	\includegraphics[scale=0.23]{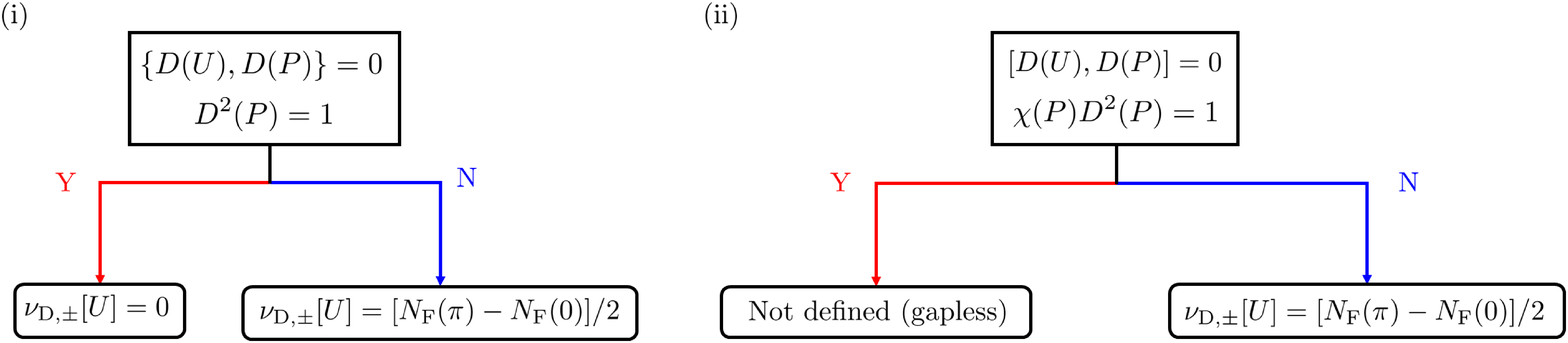}
	\caption{Fermi-surface criterion for type C, where surface symmetry satisfies $[D(U),h(k_{\perp},\bm k_{\parallel})]=0$.
	``Parity"  symmetry, which satisfies $D^{\dagger}(P)h(-k_{\perp},\bm k_{\parallel})D(P)=h(k_{\perp},\bm k_{\parallel})$, in the eigenspaces is (ii) preserved $[D(U), D(P)]=0$ or (i) not $\{D(U), D(P)\}=0$.
	Apply (i) first. If (i) is not fulfilled, then apply (ii). 
	When the system has no ``parity" symmetry, the $\mathbb Z_2$ invariant is given by $\nu_{{\rm D}, \pm}[U] = [N_{\mathrm{F}}(\pi) - N_{\mathrm{F}}(0)]/2$.
	}
	\label{TRSsymmorphic Fig}
\end{figure*}
In the following, we apply the Fermi surface criterion to the surface symmetry-protected 1D topological invariants. 
We first focus on type C, which is realized when $(\chi(U), D^2(U))=(-1,-1)$. As shown in Table \ref{symmetryclass}, the two eigenspaces of $U$ belong to class D and host the $\mathbb{Z}_2$ invariant [Eq.~(\ref{eq:nupmD})]. Following Sec.~\ref{Without time-reversal symmetry}, Eq.~(\ref{eq:nupmD}) can be reduced to 
\begin{align}
 \nu_{\text{D},\pm}[U] =N_{\mathrm{F}}^\pm(\pi) - N_{\mathrm{F}}^\pm(0) \mod 2, \label{eq:FSC_typeC}
\end{align}
where $N^{\pm}_{\rm F}$ are the number of occupied states in $H_{\pm}$. Since TRS relates $N^+_{\rm F}$ with $N^-_{\rm F}$, Eq.~(\ref{eq:FSC_typeC}) is further simplified as
\begin{align}
 \nu_{\text{D},+}[U]  = \nu_{\text{D},-}[U] = \frac{N_{\mathrm{F}}(\pi) - N_{\mathrm{F}}(0)}{2} \mod 2. \label{NF2}
\end{align}
Therefore, the  $\mathbb{Z}_2$ invariant for type C is nontrivial when the difference between the number of occupied states at $k=0$ and $\pi$ is odd in a 1D subsystem.

In addition, the even-``parity'' condition discussed in Sec.~\ref{parity} is also applied to $\nu_{\rm D, \pm}[U]$ when $[D(P), D(U)]=0$ and $\chi(P)D^2(P)=1$. 
Once the condition is satisfied, a superconducting state belongs to either a gapless phase or a topologically trivial gapped phase. 
The latter (former) phase is realized if the system has (non)degenerate Fermi surface in the $\tilde{D}(U)=\pm i$ eigenspace.
On the other hand, when $D(P)$ is an ``inversion'' symmetry and satisfies $\{D(U),D(P)\}=0$, a superconducting state always belongs to a topologically trivial gapped phase due to the Kramers degeneracy in the eigenspaces ensured by PT (``inversion'' followed by time reversal) symmetry.

We comment on the case that Eq.~(\ref{NF2}) is reduced to the number of the Fermi surfaces:
\begin{align}
 \nu_{\text{D}, \pm}[U]  = \#\mathrm{FS}/2 \mod 2,
\end{align}
which can be realized when energy bands that belong to different eigenspaces of $D(U)$ are degenerate at arbitrary $k$, i.e., the bands are doubly degenerate. The twofold degeneracy arises when systems host an ``inversion'' symmetry that satisfies $[D(U), D(P)]=0$ and $\chi(P)=-1$ or an additional surface symmetry ($U'$) that satisfies $\{D(U), D(U')\}=0$.
The flowchart for type C is summarized in Fig. \ref{TRSsymmorphic Fig}.

\subsubsection{Type D}
\label{glideZ2}
Next, we consider the $\mathbb{Z}_2$ invariant for type D, which is realized when $D(U)$ satisfies $(\chi(U), D^2(U))=(1,1)$. 
Since the eigenspaces of $\tilde{D}(U)$ belong to class DIII, the Fermi-surface criterion can be constructed in a similar way to the discussion in Sec.~\ref{With time-reversal and inversion symmetries}. 
Thus, when each eigenspace preserves ``inversion" symmetry ($P$), i.e., $\qty[D(P), D(U)]=0$ and $D^2(P)=1$, Eq. (\ref{eq:nupmDIII}) is rewritten by the number of Fermi surfaces in each eigenspace:
\begin{align}
\nu_{\text{DIII},\pm}[U] = 
 \begin{cases}
 0 & \mathrm{for} \ \chi(P)=1, \\
 \displaystyle\frac{\#\text{FS}_{\pm}}{2} 
   & \mathrm{for} \ \chi(P)=-1,
 \end{cases}
\label{Uglide}
\end{align}
with modulo 2, where $\#\mathrm{FS}_\pm$ denotes the number of Fermi surfaces in the $\tilde{D}(U)=\pm 1$ eigenspace.  
Note that $\nu_{\text{DIII},+}[U]$ and $\nu_{\text{DIII},-}[U]$ are independent hence the topological phase is characterized by $\mathbb{Z} _2 \oplus \mathbb{Z} _2$.  

\begin{figure}
	\centering
	\includegraphics[scale=0.25]{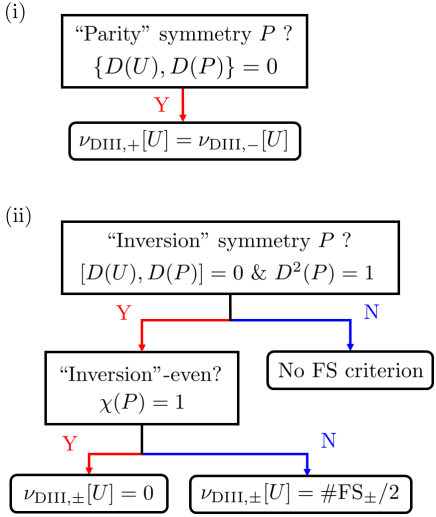}
	\caption{Fermi-surface criterion for type D with ``parity'' symmetry [$D^\dag(P) h(-k_\perp, \boldsymbol{k}_\parallel) D(P) = h(k_\perp, \boldsymbol{k}_\parallel)$]. 
	Apply criterion (i) first, then (ii). 
	If (i) is satisfied, the Fermi-surface criterion $\nu_{{\rm DIII}, \pm}[U] = \# \mathrm{FS}_\pm/2$ at the bottom of the figure reduces to $\nu_{{\rm DIII}, \pm}[U] = [N_{\mathrm F}(\pi) - N_{\mathrm{F}}(0)]/4$. 
	Furthermore if $P$ in (i) is ``inversion'' symmetry $D^2(P)=1$, the $\mathbb{Z}_2$ invariant is related to the number of Fermi surfaces $\nu_{\mathrm{DIII}, \pm}[U] = \#\mathrm{FS}_+/2 = \#\mathrm{FS}_-/2 = \# \mathrm{FS}/4$.
	}
	\label{glideZ2 Fig1}
\end{figure}

Equation (\ref{Uglide}) tells us that two $\mathbb{Z}_2$ invariants are both nontrivial when $N_{\mathrm{F}}^+(k_{\Gamma}) = N_{\mathrm{F}}^-(k_{\Gamma})$. 
Such a situation is important because Eq.~(\ref{Uglide}) manifests double Majorana Kramers pairs ($\nu_{{\rm DIII}, +}[U] = \nu_{{\rm DIII}, -}[U] = 1$), leading to a quadrupole-shaped magnetic response~\cite{yamazaki07939}.
In the following, we show that when a ``parity'' symmetry $P' (\neq P)$ or a surface symmetry $U' (\neq U)$ exists additionally, the fourfold degeneracy emerges at $k_{\Gamma}$ and satisfies $N_{\mathrm{F}}^+(k_{\Gamma}) = N_{\mathrm{F}}^-(k_{\Gamma})$. Note that the following discussion does not depend on the commutation relation between $P$ and $P'$ or $P$ and $U'$.
For an additional ``parity'' symmetry,  the fourfold degeneracy emerges when $\qty{D(P'), D(U)} = 0$, which yields $N_{\mathrm{F}}(k_\Gamma) = 2N_{\mathrm{F}}^+(k_\Gamma) = 2N_{\mathrm{F}}^-(k_\Gamma)$. Thus, Eq.~(\ref{Uglide}) is recast into
\begin{align}
\nu_{\text{DIII},+}[U]=\nu_{\text{DIII},-}[U] = \frac{N_{\mathrm{F}}(\pi) - N_{\mathrm{F}}(0)}{4} \mod 2.
\label{NF4}
\end{align}
In addition, if $P'$ is an ``inversion'' symmetry, Eq.~(\ref{NF4}) is related to the number of Fermi surfaces:
\begin{align}
 \nu_{\text{DIII},\pm}[U] = \#\mathrm{FS}/4 \mod 2, \label{eq:FSC_typeD}
\end{align}
where the fourfold degeneracy are enforced at arbitrary $k$ by PT symmetry $D(P')\Theta$.
The flowchart for type D with an additional ``parity'' symmetry is shown in Fig. \ref{glideZ2 Fig1}.

\begin{figure}
	\centering
	\includegraphics[scale=0.25]{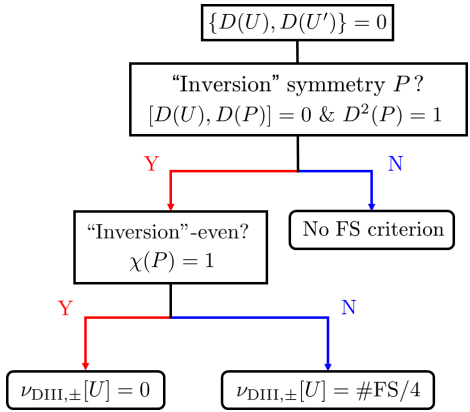}
	\caption{Fermi-surface criterion for type D with a ``parity" ($P$) and an additional surface ($U'$) symmetries.
	}
	\label{glideZ2 Fig2}
\end{figure}

Similarly, the fourfold degeneracy also emerges when an additional surface symmetry ($U'$) exists and satisfies $\{D(U), D(U')\}=0$. Since the eigenspaces of $D(U)$ are exchanged under $D(U')$, the numbers of occupied states in eigenspaces of $\tilde{D}(U)$ are the same, so we obtain Eq.~(\ref{eq:FSC_typeD}).  
The flowchart  for type D with an additional surface symmetry is shown in Fig. \ref{glideZ2 Fig2}.

\subsection{Demonstration of Fermi-surface criterion for types C and D}
\label{demo}

In this section, we demonstrate the Fermi-surface criterion of types C and D under $P222$ (No. 16) and $P2_{1}/c$ (No. 14) symmetries. 
Another example is shown in Appendix~\ref{pmma}, where we demonstrate the magnetic responses of types B and D $(\# \text{MKP}=2)$ for $Pmma$ (No. 51).

\subsubsection{Type C for \texorpdfstring{$P222$}{P222}}
\label{typeC}

Space group $P222$ consists of three rotations 
\begin{align}
	\{C_2(x) | \boldsymbol{0}\},
	\ 
	\{C_2(y) | \boldsymbol{0}\}, 
	\
	\{C_2(z) | \boldsymbol{0}\},	
\end{align}
along the $x$, $y$, and $z$ axes, respectively.
We focus on a Majorana Kramers pair on the $(xy)$ surface for $k_x=k_y=0$ with a pair potential of $B_3$ irrep of $D_2$. 
The three rotations are classified into one surface-symmetry ($U$) and two ``parity" ($P_1$ and $P_2$) symmetries as
\begin{align}
	U = \qty{C_2(z) | \bm 0},
	\ 
	P_1 = \qty{C_2(x) | \bm 0},
	\
	P_2 = \qty{C_2(y) | \bm 0}.
\end{align}
The squares of these symmetry operations equal the $2\pi$ rotation $\{{}^d \hspace{-0.4ex} E | \boldsymbol{0}\}$ and are represented by $-1$. 
They anticommute with each other. 
The $B_3$ irrep is defined by 
\begin{align}
	\chi(U)=-1, 
	\
	\chi(P_1)=1,
	\
	\chi(P_2)=-1.
\end{align}
Applying the Fermi-surface criterion for type C [$\chi(U) = D^2(U) = -1$] given in Fig. \ref{TRSsymmorphic Fig},
we find that the $\mathbb Z_2$ invariant $\nu_{\mathrm D, \pm}[U]$ protected by the twofold rotation is determined to be
\begin{align}
	\nu_{\text{D}, \pm}[U]  
	=
	\frac{N_{\mathrm{F}}(0,0, \pi) - N_{\mathrm{F}}(0,0,0)}{2} \mod 2. 
\end{align}
The energy gap of a Majorana Kramers pair corresponding to $\nu_{\text{D},\pm}[U]$ is given by 
\begin{align}
 E_{\mathrm{M}}(\boldsymbol{B})
 &
	= B_x \qty(\rho_x + \rho_{xzz} B_z^2) 
	+ B_y \qty(\rho_y + \rho_{yzz} B_z^2) 
	\nonumber\\ & \quad
	+ \order{B^5}, 
\end{align}
with satisfying $E_{\mathrm{M}}(B_x, B_y, B_z) = -E_{\mathrm{M}}(-B_x, -B_y, B_z)$ from Table \ref{fivetype}. 

\subsubsection{Type D for \texorpdfstring{$P2_1/c$}{P21/c}}
\label{typeD}

Next, we focus on Majorana Kramers pairs on the (100) surface in nonsymmorphic space group $P2_1/c$, which consists of twofold screw axis along $\boldsymbol{b}$, $c$-glide plane, and spatial inversion.
The glide plane preserves the (100) plane while the screw axis or inversion inverts it. 
The surface ($U$) and parity ($P_1$, $P_2$) symmetries are assigned to be 
\begin{align}
	U &= \{ \sigma(010) | \boldsymbol{b}/2 + \boldsymbol{c}/2 \},
	\\
	P_1 &= \qty{C_2(010) | \bm b/2 + \bm c/2},
	\
	P_2 = \qty{I | \bm 0}.
\end{align}
The products satisfy 
\begin{align}
	U^2 &= \qty{ {}^d \hspace{-0.4ex} E | \boldsymbol{c} },
	\
	P_1^2 = \qty{{}^d \hspace{-0.4ex} E | \boldsymbol{b}},
	\
	P_2^2 = \qty{E | \boldsymbol{0}},
	\\
	U P_1 &= \qty{E | -\boldsymbol{b}+\boldsymbol{c}} P_2 U,
	\
	U P_2 = \qty{E | \boldsymbol{b} + \boldsymbol{c}} P_2 U,
\end{align}
and these are represented by 
\begin{align}
	D^2(U) &= D(U^2) = -e^{-i \boldsymbol{k}_\Gamma \cdot \boldsymbol{c}},
	\\
	D^2(P_1) &= D(P_1^2) = -e^{-i \boldsymbol{k}_\Gamma \cdot \boldsymbol{b}},
	\\
	D^2(P_2) &= D(P_2^2) = 1,
	\\
	D(U) D(P_1) &= e^{-i \boldsymbol{k}_\Gamma \cdot (-\boldsymbol{b} + \boldsymbol{c})} D(P_1) D(U),
	\\
	D(U) D(P_2) &= e^{-i \boldsymbol{k}_\Gamma \cdot (\boldsymbol{b} + \boldsymbol{c})} D(P_2) D(U).
\end{align}

We consider the $B_{u}$ irrep of point group $C_{2h}$, which is defined by
$\chi(U)=1$, $\chi(P_1)=-1$, and $\chi(P_2)=-1$, 
as a representative. 
The surface symmetry satisfies $\chi(U) = D^2(U) =1$ for $\boldsymbol{k}_{\Gamma} \cdot \boldsymbol{c} = \pi$.
Applying the Fermi-surface criterion for type D [$\chi(U)=D^2(U)=1$] in Fig. \ref{glideZ2 Fig1}, 
we find that the $\mathbb Z_2$ invariant for $\boldsymbol{k}_{\Gamma} \cdot \boldsymbol{b} = \boldsymbol{k}_{\Gamma} \cdot \boldsymbol{c} = \pi$ is determined to be
\begin{align}
	\nu_{\text{DIII}, \pm}[U]  
	= \#\mathrm{FS}_{\pm}/2 \mod 2,
\end{align}
because of $[D(U), D(P)]=0$, $D^2(P)=1$, and $\chi(P)=-1$ for both $P = P_1$ and $P_2$. 
Note that $\nu_{{\rm DIII}, +}[U]$ and $\nu_{{\rm DIII}, -}[U]$ are independent since the system has no ``parity'' symmetry satisfying condition (i). 

The energy gap of a single Majorana Kramers pair [$(\nu_{\text{DIII},+}[U], \nu_{\mathrm{DIII}, -}[U]) = (1,0)$ or $(0,1)$] is given by 
\begin{align}
	E_{\mathrm{M}}(\boldsymbol{B}) 
	&= B_b 
	\qty(\rho_b + \rho_{baa} B_a^2 + \rho_{bcc} B_c^2 + \rho_{bac} B_a B_c) 
	\nonumber\\ & \quad
	 + \order{B^5}, 
\end{align}
with $f(B_a \hat{\boldsymbol{a}} + B_b \hat{\boldsymbol{b}} + B_c \hat{\boldsymbol{c}}) = f(-B_a \hat{\boldsymbol{a}} + B_b \hat{\boldsymbol{b}} - B_c \hat{\boldsymbol{c}})$
from Table~\ref{fivetype}. 
That of double Majorana Kramers pairs [$(\nu_{\text{DIII},+}[U], \nu_{\text{DIII},-}[U])=(1,1)$] is, on the other hand, given by 
\begin{align}
	E_{\text{M}}(\boldsymbol{B}) \sim \sqrt{\sum_{ij} \rho_{ij} B_i B_j} - \sqrt{\sum_{ij} \rho'_{ij} B_i B_j},
\end{align}
from Eq.~(\ref{eq:general_form_double}).

\section{Application to topological nonsymmorphic crystalline superconducting states in UCoGe}
\label{AppUCoGe} 

This section shows the magnetic response of two Majorana Kramers pairs on UCoGe, which has been proposed as a candidate time-reversal-invariant topological superconductor at high pressure \cite{Hassinger073703, Slooten097003, Bastien125110, Manago020506,Cheung134516,Mineev104501} with the space group $Pnma$ \cite{CANEPA1996225}. 
This material is a ferromagnetic superconductor at ambient pressure \cite{Huy067006, Aoki061011}, and an experimental result \cite{Hattori066403} suggests that the ferromagnetic superconducting state has $A_u$ symmetry of $C_{2h}$, 
which deforms into either $A_u$ or $B_{1u}$ symmetry of $D_{2h}$ at high pressure \cite{Daido227001}.

The $(010)$ and $(0\bar{1}1)$ surfaces have only glide-plane symmetry represented by $G_a$ and $G_n$, respectively.
On the $(010)$ surface, there is no Majorana Kramers pair protected by the $a$-glide plane ($G_a$) for the following reasons. 
UCoGe has no Fermi surface for $k_x=0$. 
For $k_x=\pi$, $D^2_{k_x=\pi, k_y, k_z}(G_a) = 1$ and $\chi(G_a) = -1$ violate a condition for the nontrivial phase for both the $A_{u}$ and $B_{1u}$ pairing.
On the $(0\bar{1}1)$ surface, on the other hand, the $B_{1u}$-pairing state hosts two Majorana Kramers pairs because of the $n$-glide-plane ($G_n$) symmetry satisfying $D_{\boldsymbol{k}}^2(G_n)= \chi(G_n) = 1$ on the Brillouin zone boundary. 
Thus, the magnetic response of Majorana Kramers pairs on UCoGe is predicted to be of type D represented by Eq.~(\ref{eq:general_form_double}). In the following, we show the magnetic response of Majorana Kramers pairs on the $(0\bar{1}1)$ surface of UCoGe with the $B_{1u}$ pairing. 

\subsection{Normal state} 

The unit cell of the system is composed of four sublattices $a1$, $a2$, $b1$, and $b2$. 
The effective model of the normal part is given by~\cite{Daido227001,Yoshida235105} 
\begin{align}
\nonumber
h(\bm{k}) &= c(\bm{k})\eta_0\sigma_0s_0 
+ t_1 \qty[\eta_0\sigma_0 + \lambda_1(k_x)] s_0 
+ \lambda_2(\bm{k})s_0 
\\ \nonumber & \quad
+ \alpha\qty[\delta_{\alpha} \sin (k_x a_1) s_y - \sin (k_y a_2) s_x]\eta_z\sigma_z \\ & \quad
+ \beta\qty[\sin(k_y a_2) s_z + \delta_{\beta}\sin(k_z a_3) s_y]\eta_z\sigma_0, \\
c(\bm{k}) &=  2t'_1\cos(k_x a_1) + 2t_2 \cos(k_y a_2) + 2t_3 \cos(k_z a_3)  
\nonumber\\ & \quad
- \mu,
\end{align}
where $s$, $\sigma$, and $\eta$ denote the Pauli matrices representing the spin, sublattice degrees of freedom $(1,2)$ and $(a,b)$, respectively. 
Here we introduce the following matrices 
\begin{align}
\lambda_1(k_x) 
&= 
\pmqty{
0 & e^{-ik_x a_1} & 0 & 0 \\
e^{ik_x a_1} & 0 & 0 & 0 \\
0 & 0 & 0 & e^{ik_x a_1} \\ 
0 & 0 & e^{-ik_x a_1} & 0},
\\
\lambda_2(\bm{k}) 
&=
\pmqty{
0 & 0 & v_1(\bm{k}) & 0 \\
0 & 0 & 0 & v_2(\bm{k}) \\
v^{*}_1(\bm{k}) & 0 & 0 & 0 \\ 
0 & v^{*}_2(\bm{k}) & 0 & 0
},
\end{align}
where $v_1(\bm{k}) = e^{-i k_x a_1}(1+e^{-ik_y a_2})(t_{ab}+e^{ik_z a_3}t'_{ab}),\ v_2(\bm{k}) = e^{ik_z a_3}(1+e^{-ik_y a_2})(t_{ab}+e^{-ik_z a_3}t'_{ab})$. The inversion $I$ and $n$-glide-plane operation $G_n$ is represented as follows.
\begin{align}
&D(I) = 
\pmqty{
0 & 0 & 1 & 0 \\
0 & 0 & 0 & 1 \\
1 & 0 & 0 & 0 \\ 
0 & 1 & 0 & 0}, \\
&D_{\bm{k}}(G_n) = (-is_x) \times
\nonumber\\& 
\pmqty{
0 & 0 & 0 & e^{i(k_x a_1 - k_y a_2)} \\
0 & 0 & e^{-ik_y a_2} & 0 \\
0 & e^{-ik_z a_3} & 0 & 0 \\ 
e^{i(k_x a_1 - k_z a_3)} & 0 & 0 & 0}.
\end{align}

\begin{figure}
 \centering
 \includegraphics[scale=0.5]{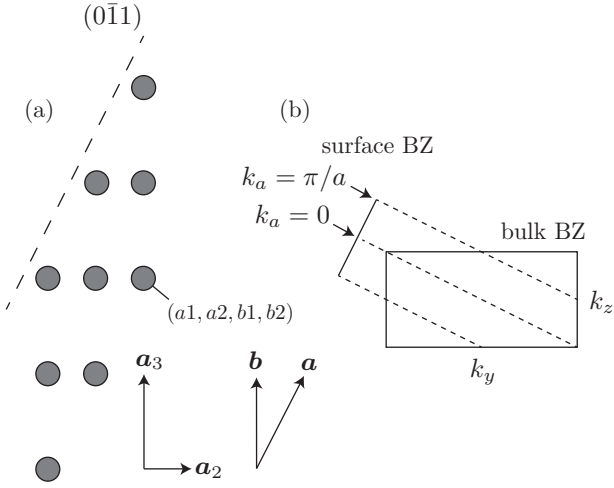}
 \caption{Geometry of the system. (a) Lattice structure on the $(yz)$ plane with the $(0 \bar 1 1)$ surfaces. Sublattice structure $(a1, a2, b1, b2)$ is regarded as an internal degree of freedom. 
 	(b) Bulk and surface Brillouin zones (BZs) for a fixed $k_x$. $a=\sqrt{a_2^2+a_3^2}$.}
 \label{fig:110}
\end{figure}

We assume that the model is defined on the primitive orthorhombic lattice spanned by $\boldsymbol a_1 = a_1 \hat{\boldsymbol{x}}$, $\boldsymbol{a}_2 = a_2 \hat{\boldsymbol{y}}$, and $\boldsymbol{a}_3 = a_3 \hat{\boldsymbol{z}}$.
Figure \ref{fig:110} shows the lattice structure with the $(0 \bar 1 1)$ surfaces on the $(yz)$ plane.
Any site in the finite-sized system terminated by the surfaces is represented by $l \boldsymbol a + m \boldsymbol{b} + n \boldsymbol{c}$ for $l$, $m$, and $n$ being integers, where we redefine translation vectors by $\boldsymbol{a} = \boldsymbol{a}_2 + \boldsymbol{a}_3$, $\boldsymbol{b} = \boldsymbol{a}_3$, and $\boldsymbol{c} = \boldsymbol{a}_1$. 
The periodic boundary condition is imposed along the $a$ and $c$ directions, while the open boundary condition along the $b$ direction. 
The quasi-one-dimensional system we calculate has the form
\begin{align}
 H = \sum_{m, m'=1}^{N_b} c^\dag_m t_{mm'}(k_a, k_c) c_{m'},
\end{align}
with annihilation operator $c_{l, m, n}$ located at $l \boldsymbol{a} + m \boldsymbol{b} + n \boldsymbol{c}$, $c_{l, m, n} = \frac{1}{\sqrt{N_a N_c}} \sum_{k_a, k_c} e^{i (k_a l a + k_c n c)} c_m(k_a, k_c)$.
The corresponding bulk Hamiltonian $H = \sum_{k_b} c^\dag(k_b) H(k_b) c(k_b)$ is obtained by the Fourier transform $c_m(k_a, k_c) = \frac{1}{\sqrt{N_b}} \sum_{k_b} e^{i k_b n b} c(k_a, k_b, k_c)$ along the $b$ direction. 
The surface zero modes on the TRIMs with $k_a a = 0$ correspond to the one-dimensional topological invariant $\nu_{{\rm D}, \pm}[G_n]$ of $H(k_b)$ while those with $k_a a = \pi$ to $\nu_{{\rm DIII}, \pm}[G_n]$. 

New coordinate $(k_a, k_b, k_c)$ is given by $k_x a_1 = k_c c$, $k_y a_2 = k_a a - k_b b$, $k_z a_3= k_c c$, where $k_a$ and $k_c$ are conserved in the system terminated with the ($0 \bar 1 1$) surface. 
\begin{figure}
	\includegraphics[scale=0.28]{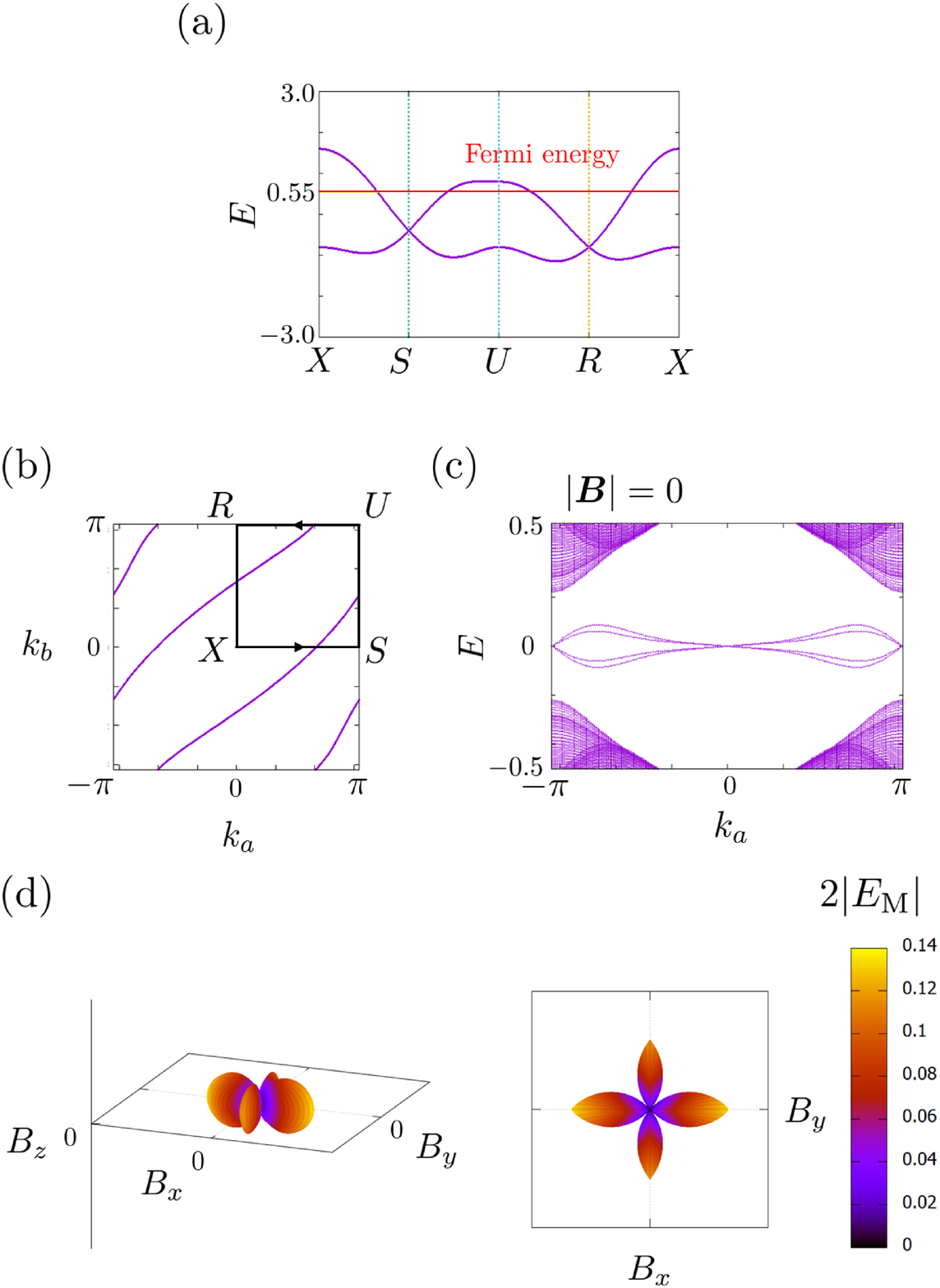}
	\caption{Energy spectra for the normal and superconducting states. (a) Energy dispersion of the normal state $h(\bm k)$. 
		(b) Fourfold-degenerate Fermi surfaces of the normal state $h(\bm k)$. 
	Note that $(k_a, k_b)$ is displayed in the orthogonal form for simplicity, although $\boldsymbol{a} \cdot \boldsymbol{b} \ne 0$. 
	(c) Energy spectrum of Eq.~(\ref{BdG22}) without a magnetic field. Two Majorana Kramers pairs are located at $k_a=\pi$. 
	They are gapped by an external magnetic field. 
	(d) The polar plot of the energy gap $|E_{\mathrm M}|$ of $H+\Delta H+H_{\mathrm{Z}}$ as a function of $\boldsymbol{B}$ with $|\vv{B}|=0.1$. The parameters are shown in Table \ref{parameterUCoGe}. 
	$\epsilon=0$ for (a) and (b). 
	The lattice constants are set to unity, $a=b=c=1$.
	For all the panels, $k_c=\pi$.}
	\label{UCoGe}
\end{figure}
Figures \ref{UCoGe}(a) and \ref{UCoGe}(b) show the band structure of the normal state $h(\bm{k})$ for $k_cc=\pi$. 
The parameters are given by \cite{Daido227001}, as shown in Table~\ref{parameterUCoGe}.
\begin{table}
	\centering
	\caption{Parameters for UCoGe taken from Ref. \cite{Daido227001}.}
	\begin{ruledtabular}
		\begin{tabular}{cccccccccccccccccccc}
			$t_1$ & $t_2$ & $t_3$ & $t_{ab}$ & $t'_{ab}$ & $\mu$ & $t'_1$ & $\alpha$ & $\delta_{\alpha}$ & $\beta$ & $\delta_{\beta}$ & $\epsilon$ & $\Delta$ & $|\vv{B}|$  \\ 
			\hline 
			$1.0$ &$0.2$ & $0.1$ & $0.5$ & $0.1$ & $0.55$ & $0.1$ & $0.3$ & $0.5$ & $0.3$ & $0.5$ & $0.015$ & $0.5$ & $0.1$ 
		\end{tabular}
	\end{ruledtabular}
	\label{parameterUCoGe}
\end{table}
Note that all the bands are fourfold degenerate at any momentum owing to PT symmetry and an additional symmetry $[h(\boldsymbol{k})|_{k_cc=\pi},\sigma_z]=0$, which satisfies $\{ \sigma_z , D_{\bm k}(G_n) \}=0$. 
The present model is oversimplified to have additional symmetry.

\subsection{Superconducting state}

We take into account a pair potential of $B_{1u}$ irrep defined by \cite{Daido227001}
\begin{align}
	\nonumber
	\Delta(\bm{k}) &= \Delta [\sin(k_x a_1) s_y + \sin(k_y a_2) s_x + \sin(k_z a_3) s_y \sigma_z] \\ \nonumber
	& = \Delta \bigl[\sin(k_aa) \cos(k_bb) s_x - \sin(k_bb) \cos(k_aa) s_x 
	\\ & \qquad + \sin(k_bb) s_y \sigma_z + \sin(k_cc) s_y \bigr].
\end{align}
In addition, to clarify the relation between the magnetic response and crystalline symmetry, we break the additional symmetry $\sigma_z$ by adding the following term
\begin{align}
\Delta H(\bm{k}) 
= \epsilon \qty[D_{\bm{k}}(G_n) + D^{\dagger}_{\bm{k}}(G_n)] 
\tau_z.
\end{align}
The present case is classified into type D, i.e., $\chi(G_n) = D^2_{\boldsymbol{k}}(G_n)|_{k_a a = k_cc=\pi} = 1$.
The eigenspace of the $n$-glide plane respects the inversion symmetry due to $D^\dag(I) D_{\boldsymbol{k}}(G_n) D(I)|_{k_aa=k_cc=\pi} = D_{\boldsymbol{k}}(G_n)|_{k_aa=k_cc=\pi, k_b \to -k_b}$.
Hence the 1D $\mathbb Z_2$ invariant on the line at $k_aa=k_cc=\pi$ is given by 
\begin{align}
\nu_{\text{DIII},\pm}[G_n] =
\frac{\# \mathrm{FS}_\pm}{2}
  \mod 2,
\end{align}
from the Fermi-surface criterion shown in Fig.~\ref{glideZ2 Fig1}.
Extracting the number of the Fermi surface from Fig.~\ref{UCoGe}(b), we find $\nu_{{\rm DIII}, \pm}[G_n] = 1$, which corresponds to the existence of double Majorana Kramers pairs on the surface, provided that the $\epsilon$ term is small. 
Note that when $\epsilon$ is zero, the same result is obtained by the Fermi-surface criterion shown in Fig.~\ref{glideZ2 Fig2} owing to the additional symmetry. 
Fourfold degenerate Fermi surfaces ($\# \mathrm{FS} = 4$) appear between the $U$ and $S$ points, as shown in Fig.~\ref{UCoGe}(b).
Therefore, the topological invariant is obtained to be $\nu_{{\rm DIII}, \pm}[G_n] = \# \mathrm{FS}/4 = 1$.

\subsection{Magnetic response}

We calculate the energy spectrum of the system with the $(0\bar{1}1)$ surface in the presence of a Zeeman field at $k_c = \pi$.  
The finite-sized BdG Hamiltonian has the form 
\begin{align}
\nonumber
 H& + \Delta H \\ \nonumber 
 &= \sum^{N_b-1}_{m=2}\sum_{\substack{\xi,\xi'= \\ a1,a2,b1,b2}}
 c_{m,\xi}^{\dagger}
  [\epsilon(k_a, k_c)]_{\xi\xi'}
 c_{m,\xi'} \\ \nonumber
&+ \sum^{N_b-2}_{m=2} \sum_{\substack{\xi,\xi'= \\ a1,a2,b1,b2}}
\qty[
c_{m,\xi}^{\dagger}
 [t_b(k_a, k_c)]_{\xi\xi'} 
	c_{m+1,\xi'}
 + \mathrm{h.c.}
 ] \\ \nonumber
&+ \sum^{N_b-3}_{m=2} \sum_{\substack{\xi,\xi'= \\ a1,a2,b1,b2}}
\qty[
c_{m,\xi}^{\dagger}
 [t_{2b}(k_a, k_c)]_{\xi\xi'} 
	c_{m+2,\xi'}
 + \mathrm{h.c.}
 ] \\ &+ H_{\text{surf}},
\label{BdG22}
\end{align}
where $H_{\text{surf}}$ denotes the surface term on $m=1$ and $m=N_b$. 
The electron operator of sublattice $\xi$ on site $m$ for momentum $k_a$ is defined by $c_{m,\xi} = (c_{m, \xi, \uparrow}, c_{m, \xi, \downarrow})^{\mathrm T}$.
The surface term $H_{\text{surf}}$ is given by
\begin{align}
\nonumber
&H_{\text{surf}} = \sum_{\xi,\xi'= a1,a2}c_{1,\xi}^{\dagger}
  [\epsilon(k_a, k_c)]_{\xi\xi'}
 c_{1,\xi'} \\ \nonumber
&+ \sum_{\xi,\xi'=a1,a2}
\qty[
c_{1,\xi}^{\dagger}
 [t_b(k_a, k_c)]_{\xi\xi'} 
	c_{2,\xi'}
 + \mathrm{h.c.}
 ]  \\ \nonumber
&+ \sum_{\xi,\xi'=a1,a2}
\qty[
c_{1,\xi}^{\dagger}
 [t_{2b}(k_a, k_c)]_{\xi\xi'} 
	c_{3,\xi'}
 + \mathrm{h.c.}
 ] \\ \nonumber
&+  \sum_{\xi,\xi'= b1,b2}c_{N_b,\xi}^{\dagger}
  [\epsilon(k_a, k_c)]_{\xi\xi'}
 c_{N_b,\xi'} \\ \nonumber
&+ \sum_{\xi,\xi'=b1,b2}
\qty[
c_{N_b-1,\xi}^{\dagger}
 [t_b(k_a, k_c)]_{\xi\xi'} 
	c_{N_b,\xi'}
 + \mathrm{h.c.}
 ]  \\ 
&+ \sum_{\xi,\xi'=b1,b2}
\qty[
c_{N_b-2,\xi}^{\dagger}
 [t_{2b}(k_a, k_c)]_{\xi\xi'} 
	c_{N_b,\xi'}
 + \mathrm{h.c.}
 ],
\end{align}
which is invariant for the $n$-glide plane
\begin{align}
 c_{m, a1} &\to - i s_x c_{m+1, b2}, 
 \\
 c_{m, a2} &\to i s_x c_{m+1, b1}, 
 \\
 c_{m, b1}
 &\to e^{-i k_a} i s_x c_{m-1, a2}, 
 \\
 c_{m, b2}
 &\to - e^{-i k_a} i s_x c_{m-1, a1}.
\end{align}
The onsite energy $\epsilon(k_a, k_c)$ and hopping $t_b(k_a, k_c), t_{2b}(k_a, k_c)$ are obtained by replacing $1 \to c^\dag_n c_n$, $\cos k_b \to \frac{1}{2} c^\dag_n c_{n+1} + \mathrm{h.c.}$, $\sin k_b \to -\frac{i}{2} c^\dag_n c_{n+1} + \mathrm{h.c.}$, $\cos 2k_b \to \frac{1}{2} c^\dag_n c_{n+2} + \mathrm{h.c.}$, and $\sin 2k_b \to -\frac{i}{2} c^\dag_n c_{n+2} + \mathrm{h.c.}$. 
A magnetic field induces the Zeeman term
\begin{align}
H_{\mathrm Z} = \sum^{N_b}_{m=1} 
\sum_{\xi=a1,a2,b1,b2}
c_{m,\xi}^{\dagger} \vv{B} \cdot \vv{s} \sigma_0 \tau_0 c_{m, \xi}.
\end{align}

The energy spectrum of the total Hamiltonian $H + \Delta H + H_{\mathrm Z}$ for the $B_{1u}$ pairing is shown in Figs.~\ref{UCoGe}(c) and \ref{UCoGe}(d). 
Without a magnetic field $|\vv B|=0$, the superconducting gap in the bulk is of the order of $0.5$.
Two Majorana Kramers pairs on the surface for $k_aa=\pi$ appear. 
Note that the energy spectrum has a small gap for $k_a=0$. 
Figure \ref{UCoGe}(d) shows the polar plot of the energy gap $|E_{\text{M}}(\vv{B})|$ of $H+\Delta H+H_{\mathrm{Z}}$ as a function of $\boldsymbol{B}$ with $|\vv{B}|=0.1$. The energy gap of Majorana Kramers pairs is of the order of $\sim 0.14$ at maximum and biaxially (quadrupolar) anisotropic in the $(0 \bar 1 1)$ plane.
This behavior is a direct consequence of double Majorana Kramers pairs protected by glide-plane symmetry.

\section{Conclusion} 
\label{conclusion}

We have provided a comprehensive understanding of the magnetic response of $\mathbb Z_2$-protected Majorana Kramers pairs emergent on the surface of TCSCs with order-two crystalline symmetries. 
Four types A, B, C, and D of the magnetic response can occur, depending on crystalline and Cooper-pair symmetries, and especially on the number of Majorana Kramers pairs.  
We have also constructed a practical procedure to diagnose electromagnetic degrees of freedom of Majorana Kramers pairs only from the Fermi-surface topology and symmetry consideration, that is, an extended Fermi-surface criterion of $\mathbb Z_2$ invariants protected by order-two symmetry for arbitrary space group including nonsymmorphic ones. 
Applying the procedure, we found a biaxially (quadrupole-shaped) magnetic response on candidate TCSC UCoGe, which would be direct evidence of the presence of double Majorana Kramers pairs protected by the glide-plane symmetry.

The predicted magnetic response can be measured in a spectroscopic way, e.g., tunneling spectroscopy on a surface of TCSC under a magnetic field or with a ferromagnet attached \cite{tanaka95, fogelstrom97, tanaka02, tanuma02, tanaka09, tamura17}. 
Gap opening in Majorana Kramers pairs on the zero energy leads to the split of zero-bias conductance peak. We have focused on Majorana Kramers pairs on TRIMs.
Due to the self-conjugate property $\gamma = \gamma^\dag$ of Majorana fermion \cite{law09, yamakage14},
they dominantly contribute to the zero-bias tunneling conductance $G=2e^2/h \sum_{\boldsymbol{k}_\parallel} T_{\boldsymbol k_\parallel}$, where $T_{\boldsymbol{k}_\parallel}$ denotes the transmission probability on the zero energy given by $T_{\boldsymbol{k}_\parallel} = 1 + A_{\boldsymbol{k}_\parallel} - B_{\boldsymbol{k}_\parallel}$ with $A_{\boldsymbol{k}_\parallel}$ and $B_{\boldsymbol{k}_\parallel}$ the Andreev and normal reflection probabilities, respectively. The resonant Andreev reflection occurs for self-conjugate states and leads to a sizable zero-bias tunneling conductance with $A_{\boldsymbol{k}_\parallel} = B_{\boldsymbol{k}_\parallel}$, i.e., $T_{\boldsymbol{k}_\parallel} = 1$ for $\boldsymbol{k_\parallel} \in \mathrm{TRIM}$. Since zero-energy states away from TRIMs do not hold the self-conjugate property, $T_{\boldsymbol{k}_\parallel}$ takes a value smaller than unity. 
Nevertheless, they may smear the splitting of the zero-bias conductance peak. Therefore, for comparison our theory to an experiment, we must quantitatively estimate the conductance spectrum attributed to Majorana Kramers pairs. 

Surface-sensitive measurements such as ferromagnetic resonance of a ferromagnet/TCSC junction \cite{inoue17, kato19} can also detect the anisotropy of Majorana Kramers pairs via the surface spin susceptibility. 
We have focused especially on the Zeeman effect from a magnetic field. The other contribution from an external magnetic field is the screening current, the Meissner effect. The coupling is well-described by the Doppler shift of energy dispersion in the weak-field regime \cite{Chirolli094515}. It is guessed that this effect plays a small role in the magnetic response of Majorana Kramers pairs since the energy on TRIMs does not change. However, it is necessary to calculate how much it affects the physical quantity, such as the density of states and magnetic susceptibility. 

In the present work, we consider only a classical magnetic field as a magnetic perturbation. 
Majorana Kramers pairs can, on the one hand, couple to magnetic degrees of freedom with quantum dynamics, such as a magnetic impurity leading to anisotropic Kondo effect~\cite{shindou10}. 
Our theory will be useful for deducing a Hamiltonian coupled to dynamics variables.
These will be studied in future work. 
Although the theory applies to limited cases in this paper, it can apply to all the space groups with any surface and irreps of pair potential. 
The complete data for 230 space groups would be more efficient for practical applications.
From a fundamental viewpoint, it is interesting to consider an exotic electromagnetic response of Majorana Kramers pairs protected by $C_3$, $C_4$, or $C_6$ symmetries, which are out of the scope of this paper.  
Moreover, the discussion in Sec.~\ref{Magnetic} implies that Majorana Kramers pairs posses electric degrees of freedom whose ranks are higher than the monopole. 
The electric operators can be decomposed into irreps of space group in a way similar to magnetic ones. 
These issues will be addressed in a future paper~\cite{kobayashi_full}.

\begin{acknowledgments}
A.Y. was supported by JSPS KAKENHI Grants Nos.~JP20K03835 and JP20H04635. 
S.K. was supported by JSPS KAKENHI Grants No.~JP19K14612 and by the CREST project (JPMJCR16F2,~JPMJCR19T2) from Japan Science and Technology Agency (JST).
\end{acknowledgments}

\appendix

\section{Fermi-surface criterion when types C and D coexist}
\label{sec:coexistence}

We consider a particular case of the Fermi-surface criterion that hosts two surface symmetries $U$ and $U'$ such that
\begin{subequations}
\label{eq:typeCandD}
\begin{align}
&
\chi(U)=D^2(U)=1, \label{eq:condtypeD}\\
&
\chi(U')=D^2(U')=-1, \label{eq:condtypeC} \\ 
&
\qty[D(U), D(U')] = 0.
\end{align}
\end{subequations}
Here, $D(U)$ and $D(U')$ satisfy the conditions for the types D and C, respectively. 
The situation is often realized for nonsymmorphic space group symmetry like $Pmma$ (No. 51), which we will discuss in detail in Appendix~\ref{pmma}. 
Since they are commutative, the Hamiltonian can be divided into four eigenspaces in terms of $\tilde{D}(U) = \pm 1$ and $\tilde{D}(U')=\pm i$, each of which belongs to class D and is classified by $\mathbb Z_2$ invariants $\nu_{\pm,\pm i}$. The $\mathbb Z_2$ invariants are related to Eqs.~(\ref{eq:nupmD}) and (\ref{eq:nupmDIII}):
\begin{align}
 &\nu_{\text{DIII},\pm}[U] = \nu_{\pm, +i} \mod 2, \label{eq:relnuDIII}
 \\
 &\nu_{\text{D},\pm}[U'] = \nu_{+, \pm i} + \nu_{-, \pm i} \mod 2, \label{eq:relnuD}
\end{align}
where $\nu_{\pm, +i} = \nu_{\pm, -i}$ due to TRS. 
Thus, whole systems are classified by $(\nu_{+, +i},\nu_{-, +i}) \in \mathbb{Z}_2 \oplus \mathbb{Z}_2$.
Note that for a single Majorana Kramers pair, Eqs.~(\ref{eq:relnuDIII}) and (\ref{eq:relnuD}) are both nontrivial, and thus, the magnetic response is determined from both Eqs.~(\ref{eq:mag_typB}) and (\ref{eq:mag_typC}), namely, $E_{\mathrm{M}}(\boldsymbol{B}) = \pm f(\boldsymbol{B})$ with
\begin{align}
 f(U\boldsymbol{B}) = f(\boldsymbol{B}),
 \
 f(U'\boldsymbol{B}) = -f(\boldsymbol{B}),
\end{align} 
which can induce a complicated magnetic response.
On the other hand, for double Majorana Kramers pairs, Eq.~(\ref{eq:relnuDIII}) remains but Eq.~(\ref{eq:relnuD}) becomes zero, so that the magnetic response is described by Eq.~(\ref{eq:general_form_double}).
Due to the fact that the four eigenspaces belong to class D, the Fermi-surface criterion [Eq.~(\ref{NF2})] is applied to each eigenspace, and $\nu_{\pm,+ i}$ is rewritten as
\begin{align}
\nu_{\pm, +i} =  \frac{N_{\mathrm F}^\pm(\pi) - N_{\mathrm{F}}^\pm(0)}{2} \mod 2,
\label{UU'}
\end{align}
where $N_{\mathrm{F}}^{\pm}(k_\Gamma)$ denotes the occupation number of the eigenspace with $\tilde D_{\bm{k}_{\Gamma}}(U)=\pm 1$.

\begin{figure}
	\centering
	\includegraphics[scale=0.25]{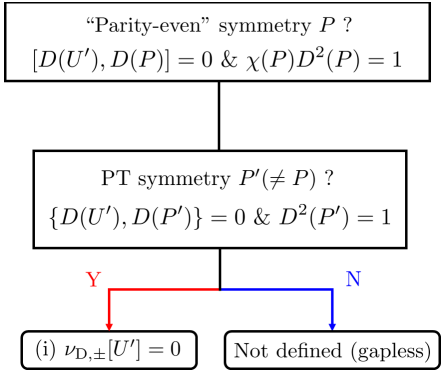}
	\caption{Fermi--surface criterion of $\nu_{\mathrm D, \pm}[U']$ for systems with two surface symmetries $U$ and $U'$ that satisfy Eq.~(\ref{eq:typeCandD}).}
	\label{With time-reversal, symmorphic, and nonsymmorphic surface symmetries Fig1}
\end{figure}

\begin{figure}
	\centering
	\includegraphics[scale=0.245]{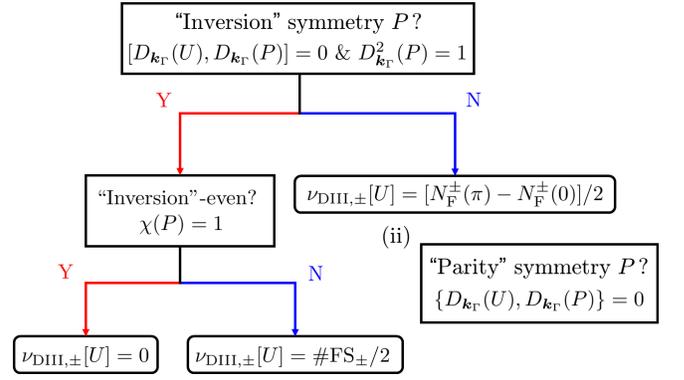}
	\caption{Fermi-surface criterion of $\nu_{\mathrm{DIII}, \pm}[U]$ for systems with two surface symmetries $U$ and $U'$ that satisfy Eq.~(\ref{eq:typeCandD}). 
	If condition (i) or (ii) in Figs.~\ref{With time-reversal, symmorphic, and nonsymmorphic surface symmetries Fig1}--\ref{With time-reversal, symmorphic, and nonsymmorphic surface symmetries Fig2} holds, then $\nu_{\text{DIII},+}[U]=\nu_{\text{DIII},-}[U]$}
	\label{With time-reversal, symmorphic, and nonsymmorphic surface symmetries Fig2}
\end{figure}

An additional ``parity'' symmetry further simplifies the Fermi--surface criterion in a similar way to types C and D discussed in Secs.~\ref{TRSsymmorphic}~and~\ref{glideZ2}.
The Fermi--surface criterion for this case is summarized in Figs.~\ref{With time-reversal, symmorphic, and nonsymmorphic surface symmetries Fig1}~and~\ref{With time-reversal, symmorphic, and nonsymmorphic surface symmetries Fig2}.

\section{Magnetic response of Majorana Kramers pairs for \texorpdfstring{$Pmma$}{Pmma}}
\label{pmma}

Here we show another example of magnetic responses of Majorana Kramers pairs on the $(xz)$ surface in space group $Pmma$ (No.~51).
The result is summarized in Table~\ref{Pmma}. 
\begin{table*}
\caption{
 Magnetic response of Majorana Kramers pairs on the $(xz)$ surface for $k_x=0$ and $k_x=\pi$ in space group $Pmma$. 
 Irreducible representation (irrep) of the superconducting pair potential in the bulk, 
 topological invariant associated with the symmetry $U_i$ which protects Majorana Kramers pairs, 
 the number of Majorana Kramers pairs (\#MKP) [$N$ denotes a nonnegative integer], 
 and the energy gap $E_{\mathrm M}(\boldsymbol{B})$ induced by a magnetic field $\boldsymbol{B}$ are shown.
 ``gap" and ``node" denote the presence and absence of superconducting gap, respectively, in the weak-pairing regime $|\Delta(k_{\mathrm F})| \ll |h(k_{\mathrm F})-\mu|$ \cite{Kobayashi180504, Sumita134512, Sumita134513, Yoshida235105} on the Fermi surface $k_\perp=k_{\mathrm F}$.
 The $\mathbb Z_2$ indices here are determined by the Fermi-surface criterion (Sec.~\ref{fermi}). 
 The surface symmetry operations are assigned to be 
 $U_1 = \qty{C_2(y) | \bm 0}$, 
 $U_2 = \qty{\sigma(xy) | \bm a/2}$, 
 and
 $U_3 = \qty{\sigma(yz) | \bm a/2}$.
}
\begin{ruledtabular}
	\begin{tabular}{lllllc}
        {$k_x=0$} & & & & & \\
		\hline
		irrep & \multicolumn{2}{l}{Topo} & \#MKP & $E_{\mathrm M}(\boldsymbol B)$ & $|\Delta| \ll |h-\mu|$
		\\
		\hline
		$A_g$ & 0 & & 0 & - & gap
		\\
		$B_{1g}$ & 0 & & 0 & - & node
		\\
		$B_{2g}$ 	& 0 & & 0 & - & node
		\\
		$B_{3g}$ 	& 0 & & 0 & - & node
		\\
		$A_u$ & $\mathbb{Z}$ & $W[U_1]$ & $N$ & $\propto B_y$ & gap
		\\
		& $\mathbb Z_2$ & $\nu_{\text{D}, \pm}[U_2] = \nu_{\text{D}, \pm}[U_3] = (\#\mathrm{FS})/2$ & & & 
		\\
		$B_{1u}$ 	& $\mathbb{Z}$ & $W[U_3]$ & $N$ & $\propto B_x$ & gap
		\\
		& $\mathbb Z_2$ & $\nu_{\text{D}, \pm}[U_1] = \nu_{\text{D}, \pm}[U_2] = (\#\mathrm{FS})/2$ & & & 
		\\
		$B_{2u}$ 	& 0 & & 0 & - & node
		\\
		$B_{3u}$ 	& $\mathbb{Z}$ & $W[U_2]$ & $N$ & $\propto B_z$ & gap
		\\
		& $\mathbb Z_2$ & $\nu_{\text{D}, \pm}[U_1] = \nu_{\text{D}, \pm}[U_3] = (\#\mathrm{FS})/2$ & & &
		\\
		\hline\hline
        {$k_x=\pi$} & & & \\
		\hline
		irrep & \multicolumn{2}{l}{Topo} & \#MKP & $E_{\mathrm M}(\boldsymbol B)$ & $|\Delta| \ll |h-\mu|$
		\\
		\hline
		$A_g$ & $2\mathbb{Z}$ & $W[U_3]$ & $2N$ & $\propto B_x$ & gap
		\\
		& $\mathbb Z_2$ & $\nu_{\text{DIII}, \pm}[U_2]$ & & &
		\\
		$B_{1g}$	& $\mathbb Z_2$ & $\nu_{\text{DIII}, \pm}[U_2]  = \qty[N_{\mathrm{F}}(\pi) - N_{\mathrm{F}}(0)]/4$
		& 0, 2 & $\sqrt{ \sum_{ij} \rho_{ij} B_i B_j} - \sqrt{ \sum_{ij} \rho'_{ij} B_i B_j}$ & node
		\\
		$B_{2g}$ 	& 0 & & 0 & - & node
		\\
		$B_{3g}$ 	& 0 & & 0 & - & node
		\\
		$A_u$ 		& 0 & & 0 & - & node
		\\
		$B_{1u}$ 	& 0 & & 0 & - & gap
		\\
		$B_{2u}$ 	& $2\mathbb Z$ & $W[U_1]$ & $2N$ & $\propto B_y$ & gap
		\\
		& $\mathbb{Z}_2$ & $\nu_{\text{DIII}, \pm}[U_2]$ & & & 
		\\
		$B_{3u}$ 	& $\mathbb Z_2$ & $\nu_{\text{DIII}, \pm}[U_2]  = \qty[N_{\mathrm{F}}(\pi) - N_{\mathrm{F}}(0)]/4$
		& 0, 2 & $\sqrt{ \sum_{ij} \rho_{ij} B_i B_j} - \sqrt{ \sum_{ij} \rho'_{ij} B_i B_j}$ & node
	\end{tabular}
\end{ruledtabular}
\label{Pmma}
\end{table*}

\subsection{General classification}
\label{General classification}

\subsubsection{Symmetry operation}

The primitive translation vector $\bm a$ is set to be along the $x$ direction. 
There are three surface-symmetry operations 
\begin{align}
 U_1 = \qty{C_2(y) | \bm 0}, 
 U_2 = \qty{\sigma(xy) | \bm a/2},
 U_3 = \qty{\sigma(yz) | \bm a/2},
\end{align}
(and identity)
which preserve the $(xz)$ surface and four ``parity" operations
\begin{align}
 P_1 = \qty{\sigma(xz) | \bm 0}, 
 P_2 = \qty{C_2(z) | \bm a/2},
 P_3 = \qty{C_2(x) | \bm a/2},
\end{align}
and inversion $I=U_iP_i$ which invert the $(xz)$ surface. 

The square of order-two symmetry operation $g^2 = \{{}^d E | 2 \boldsymbol{\tau}_g\}$, except for inversion, is represented as 
\begin{align}
 D_{\boldsymbol{k}}(g^2) = D_{\boldsymbol{k}}^2(g) = -e^{-i \boldsymbol{k} \cdot 2 \boldsymbol{\tau}_g},
\end{align}
on the $g$-invariant momentum $\boldsymbol{k}$. 
Inversion $I$ is squared to be $D^2_{\boldsymbol{k}_\Gamma}(I) = 1$ on $\boldsymbol{k}_\Gamma$ being a TRIM.
The commutation relations $g_1 g_2 = \qty{{}^dE | R_{g_1} \boldsymbol{\tau}_{g_2} + \boldsymbol{\tau}_{g_1} - R_{g_2} \boldsymbol{\tau}_{g_1} - \boldsymbol{\tau}_{g_2} }  g_2 g_1$ and $g I = \qty{E | 2\boldsymbol{\tau}_{g}}  I g$ are represented as
\begin{align}
 &
 D_{\boldsymbol{k}}(g_1) D_{\boldsymbol{k}}(g_2)
 \nonumber\\ &
 =
 - e^{-i \boldsymbol{k} \cdot (R_{g_1} \boldsymbol{\tau}_{g_2} + \boldsymbol{\tau}_{g_1} - R_{g_2} \boldsymbol{\tau}_{g_1} - \boldsymbol{\tau}_{g_2})}
 D_{\boldsymbol{k}}(g_2) D_{\boldsymbol{k}}(g_1),
 \\ &
 D_{\boldsymbol{k}}(g) D_{\boldsymbol{k}}(I)
 = e^{-i \boldsymbol{k} \cdot 2 \boldsymbol{\tau}_g} 
 D_{\boldsymbol{k}}(I)
  D_{\boldsymbol{k}}(g),
\end{align}
for order-two symmetry operations on TRIMs. 
These relations for $Pmma$ are listed in Table \ref{table_mwn2}.

\begin{table}
	\caption{Representation matrices $D_{\boldsymbol{k}}(g)$ and characters $\chi_\Gamma(g)$ for the irrep $\Gamma$ of the space group $Pmma$, for $g=U_i$ and $P_j$. 
		$\eta_{\boldsymbol{k}}(g_1, g_2)$ denotes the commutation relation between the representation matrices, which is defined by $D_{\boldsymbol{k}}(g_1) D_{\boldsymbol{k}}(g_2) = \eta_{\boldsymbol{k}}(g_1, g_2) D_{\boldsymbol{k}}(g_2) D_{\boldsymbol{k}}(g_1)$.
		The surface- and ``parity"-symmetry operations are assigned to be 
		$U_1 = \{C_{2}(y) | \boldsymbol 0 \}$,
		$U_2 = \{ \sigma(xy) | \boldsymbol{a}/2 \}$, 
		$U_3 = \{ \sigma(yz) | \boldsymbol{a}/2 \}$,
		$P_1 = \{ \sigma(xz) | \boldsymbol{0} \}$,
		$P_2 = \{ C_{2}(z) | \boldsymbol{a}/2 \}$,
		$P_3 = \{ C_{2}(x) | \boldsymbol{a}/2 \}$,
		and the inversion $I$.
	}
	\begin{ruledtabular}
		\begin{tabular}{cccccccc}
			& $U_1$ & $U_2$ & $U_3$ & $I$ & $P_1$ & $P_2$ & $P_3$
			\\
			\hline
			$D_{\boldsymbol{k}}^2(g)$ & $-$ & $-e^{-i k_x}$ & $-$ & $+$ & $-$ & $-$ & $-e^{-i k_x}$
			\\
			$\chi_{A_g}(g)$ & $+$ & $+$ & $+$ & $+$ & $+$ & $+$ & $+$
			\\
			$\chi_{B_{1g}}(g)$ & $-$ & $+$ & $-$ & $+$ & $-$ & $+$ & $-$
			\\
			$\chi_{B_{2g}}(g)$ & $+$ & $-$ & $-$ & $+$ & $+$ & $-$ & $-$
			\\
			$\chi_{B_{3g}}(g)$ & $-$ & $-$ & $+$ & $+$ & $-$ & $-$ & $+$
			\\
			$\chi_{A_u}(g)$ & $+$ & $-$ & $-$ & $-$ & $-$ & $+$ & $+$
			\\
			$\chi_{B_{1u}}(g)$ & $-$ & $-$ & $+$ & $-$ & $+$ & $+$ & $-$
			\\
			$\chi_{B_{2u}}(g)$ & $+$ & $+$ & $+$ & $-$ & $-$ & $-$ & $-$
			\\
			$\chi_{B_{3u}}(g)$ & $-$ & $+$ & $-$ & $-$ & $+$ & $-$ & $+$
			\\
			$\eta_{\boldsymbol{k}}(U_1, g)$ & 1 & $-e^{i k_x}$ & $-e^{i k_x}$ & 1 & 1 & $-e^{i k_x}$ & $-e^{i k_x}$
			\\
			$\eta_{\boldsymbol{k}}(U_2, g)$ & $-e^{i k_x}$ & 1 & $-e^{i k_x}$ & $e^{ik_x}$ & $-1$ & $e^{i k_x}$ & $-1$
			\\
			$\eta_{\boldsymbol{k}}(U_3, g)$ & $-e^{i k_x}$ & $-e^{i k_x}$ & 1 & $e^{ik_x}$ & $-1$ & $-1$ & $e^{i k_x}$
		\end{tabular}
	\end{ruledtabular}
	\label{table_mwn2}
\end{table}

\subsubsection{\texorpdfstring{$\mathbb Z$}{Z} invariant}

We use the irreps of point group $D_{2h}$, which is compatible with space group $Pmma$, to classify the pair potentials. 
A necessary condition for type--B Majorana Kramers pairs, $W[U_i] \ne 0$, has been obtained \cite{xiong17} and summarized as
\begin{align}
\nonumber
D_{\boldsymbol{k}}^2(U_i) &= -1,
\
\chi(U_i) = 1,
\\
\chi(U_j) &= \eta_{\boldsymbol{k}}(U_i, U_j),
\
\chi(P_j) = -\eta_{\boldsymbol{k}}(U_i, P_j),
\label{condition}
\end{align}
where $\eta_{\boldsymbol{k}}(g_1, g_2)$ denotes the commutation relation between $D_{\boldsymbol{k}}(g_1)$ and $D_{\boldsymbol{k}}(g_2)$, defined by $D_{\boldsymbol{k}}(g_1) D_{\boldsymbol{k}}(g_2) = \eta_{\boldsymbol{k}}(g_1, g_2) D_{\boldsymbol{k}}(g_2)D_{\boldsymbol{k}}(g_1)$.

For $k_x=0$, odd--parity pairings except for $B_{2u}$ satisfy the above condition, i.e., $A_u$, $B_{1u}$, and $B_{3u}$ pairings can host Majorana Kramers pairs protected by the magnetic winding numbers $W[U_1]$, $W[U_3]$, and $W[U_2]$, respectively. 
The character $\chi_\Gamma(g)$ of $g$ for irrep $\Gamma$ is shown in Table \ref{table_mwn2}. 
Since the commutation relations for $k_x=\pi$ change from those for $k_x=0$, as shown in Table \ref{table_mwn2}, $A_g$ and $B_{2u}$ pairings can take a nontrivial value of $W[U_3]$ and $W[U_1]$, respectively. 
The corresponding magnetic responses are also summarized in Table \ref{Pmma}.

\subsubsection{\texorpdfstring{$\mathbb Z_2$}{Z2} invariant}

The $\mathbb Z_2$ invariants $\nu_{\mathrm D, \pm}[U]$ and $\nu_{\mathrm{DIII}, \pm}[U]$ can be nontrivial for types C [$(\chi(U), D^2(U))=(-1, -1)$] and D [$(\chi(U), D^2(U)) = (1, 1)$], respectively. 
The possible nontrivial case for $A_g$ pairing is only $\nu_{\mathrm{DIII}, \pm}[U_2]$ on $k_x=\pi$. 
Since all the parities exchange the $U_2$ eigenspaces, the invariant $\mathbb Z_2 \oplus \mathbb Z_2$ reduces to $\mathbb Z_2$, $\nu_{\mathrm{DIII},+}[U_2] = \nu_{{\rm DIII}, -}[U_2]$, but no Fermi--surface criterion applies to them, as shown in Fig.~\ref{glideZ2 Fig1}. 
As a result, the magnetic winding number $W[U_3]$, which is shown above, coexists $\nu_{{\rm DIII}, +}[U_2] = \nu_{{\rm DIII}, -}[U_2]$. 
The former invariant determines the magnetic response, $E_{\mathrm M}(\boldsymbol{B}) \propto B_x$, and must be an even number $W[U_3] \in 2 \mathbb Z$ owing to the latter condition. 

For the $B_{1g}$ pairing, on the other hand, the $\mathbb Z_2$ invariant $\nu_{\text{D}, \pm}[U_1]$ cannot be defined (gapless) or zero, since the even--``parity" condition, $[D(I), D(U_1)]=0$ and $\chi(I) D^2(I)=1$, is fulfilled (see Fig.~\ref{TRSsymmorphic Fig}). 
Such a symmetry-protected gap node is indicated in the last column of Table.~\ref{Pmma}.
An alternative and systematic analysis of the nodes is shown in Appendix~\ref{node}.
Topological invariants and magnetic responses for the other irreps are deduced in a similar way and summarized in Table~\ref{Pmma}.

\subsection{Toy model}
\label{toy_model}

\begin{figure}
	\centering
	\includegraphics{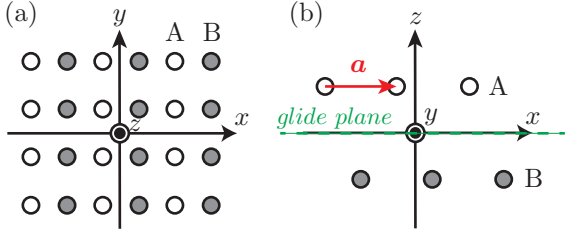}
	\caption{Lattice structure of the toy model. Top (a) and side (b) views. 
		$\boldsymbol{a}$ is the primitive translational vector along the $x$ axis. 
		There are two sublattices denoted by A (open circle) and B (closed circle). 
		The $(xy)$ plane (the dashed green line) is the glide plane.}
	\label{model}
\end{figure}

Next we verify the above general result by examining a toy model on a layered 2D lattice with two sublattices which has the glide plane on the $(xz)$ surface \cite{wang-liu}, as shown in Fig.~\ref{model}. 

\subsubsection{Normal state}

The normal part $h(\vv k)$ of the BdG Hamiltonian reads 
\begin{align}
\nonumber
h(\vv k) &= c(\vv k)\sigma_0 s_0 + t_3 \cos(k_x/2) \sigma_1(k_x) s_0 
\\ & \quad + (\lambda_1 s_x \sin k_y + \lambda_2  s_y \sin k_x) \sigma_3,
\label{normalpart}
\\ \nonumber
c(\vv k) &= m_0 + t_1 \cos k_x + t_2 \cos k_y,
\end{align}
where the lattice constant is set to 1.
$s$ and $\sigma$ denote the Pauli matrices representing the spin and layer degrees of freedom (A and B), respectively. 
Here we introduce the modified Pauli matrices
\begin{align}
\sigma_1(k_x) 
&= 
\pmqty{
0 & e^{ik_x/2} \\
e^{-ik_x/2} & 0
},
\\
\sigma_2(k_x) 
&=
\pmqty{
0 & -ie^{ik_x/2} \\
ie^{-ik_x/2} & 0
},
\end{align}
and $\sigma_3=\text{diag}(1,-1)$,
which satisfy $\{\sigma_i(k_x), \sigma_j(k_x)\} = 2 \delta_{ij}$. 
$h(\boldsymbol{k})$ has to satisfy the periodicity $h(\boldsymbol{k}) = h(\boldsymbol{k} + \boldsymbol{G})$, where $\boldsymbol{G}$ denotes a reciprocal lattice vector. 
Note that $\sigma_1(k_x)$ and $\sigma_2(k_x)$ are anti-periodic; $\sigma_1(k_x) = -\sigma_1(k_x+2\pi)$ and $\sigma_2(k_x) = -\sigma_2(k_x+2\pi)$.
Symmetry operations of $Pmma$ are represented as follows.
\begin{align}
 D_{\boldsymbol{k}}(\qty{C_2(y)|\boldsymbol{0}}) &= i \sigma_1 s_y,
 \\
 D_{\boldsymbol{k}}(\qty{\sigma(xy)|\boldsymbol{a}/2}) 
 &= i
 \pmqty{0 & e^{-i k_x}
\\
 1 & 0 
} 
 s_z,
 \\
 D_{\boldsymbol{k}}(\qty{\sigma(yz)|\boldsymbol{a}/2}) 
 &= 
 i
 \pmqty{e^{i k_x} & 0
\\
 0 & 1 
} s_x,
\\
 D_{\boldsymbol{k}}(\qty{\sigma(xz)|\boldsymbol{0}})
 &= i \sigma_0 s_y,
 \\
 D_{\boldsymbol{k}}(\qty{C_2(z)|\boldsymbol{a}/2})
 &=
 i \pmqty{e^{ik_x} & 0
\\
 0 & 1 
} s_z,
\\
 D_{\boldsymbol{k}}(\qty{C_2(x)|\boldsymbol{a}/2})
 &=
 i
  \pmqty{
 0 & e^{-ik_x}
 \\
  1 & 0
} s_x.
\end{align}

\begin{figure}
	\centering
	\includegraphics[scale=0.30]{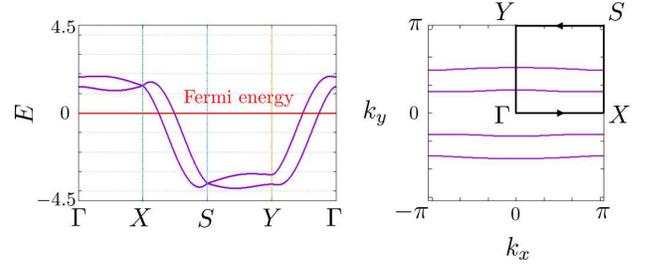}
	\caption{Energy dispersion (left) and Fermi surface (right) of the normal state $h(\vv k)$. }
	\label{band}
\end{figure}

The band structure of the normal state $h(\vv k)$ is shown in Fig. \ref{band}. 
\begin{table}
	\centering
	\caption{Parameters taken for Eq. (\ref{normalpart}).}
	\begin{ruledtabular}
		\begin{tabular}{cccccccccc}
			 $m_0$ & $t_1$ & $t_2$ & $t_3$ & $\lambda_1$ & $\lambda_2$ & $\Delta$ & $\Delta'$ & $|\vv{B}|$  \\ 
			\hline 
			 $-1.0$ & $0.1$ & $2.5$ & $0.25$ & $-1.0$ & $0.3$ & $1.5$ & $0.5$ & $0.2$
		\end{tabular}
	\end{ruledtabular}
	\label{parameter}
\end{table}
The parameters are set to those shown in Table \ref{parameter}.
All the bands are twofold degenerate at any momentum owing to inversion and time-reversal symmetries. 
Particularly, the bands are fourfold degenerate at the $X$ and $S$ points. 
Therefore the number of Fermi surfaces is a multiple of 4 between the $X$ and $S$ points.

\subsubsection{Superconducting state}

Next, we consider time-reversal-invariant superconducting pair potential $\Delta_\Gamma(\boldsymbol{k})$, where $\Gamma$ labels irreps of $D_{2h}$.
The BdG Hamiltonian has the form
\begin{align}
 H(\boldsymbol{k}) = \qty[h(\boldsymbol{k})-\mu] \tau_z + \Delta_\Gamma(\boldsymbol{k}) \tau_x.
 \label{BDGD}
\end{align}
The possible pair potentials are listed in Table \ref{pairpotential}.
\begin{table}
	\caption{Time-reversal-invariant pair potential $\Delta_\Gamma(\bm k)$ for irrep $\Gamma$ that is nonzero for $k_x=0$ or $\pi$ in the $Pmma$ model.}
	\begin{ruledtabular}
	\begin{tabular}{ll}
		irrep & $\Delta_\Gamma(\bm k)$
		\\ \hline 
		$A_g$ & 
		$\Delta \sigma_3 s_x \sin k_y$
		\\ 
		$B_{1g}$ & $\Delta \sigma_3 s_y \sin{k_y} + \Delta' \sigma_1(k_x) s_0 \sin(k_x/2) \sin k_y$ 
		\\ 
		$B_{2g}$ & $\Delta \sigma_3 s_z \sin k_y + \Delta' \sigma_2(k_x)s_0 \sin(k_x/2)$  
		\\ 
		$B_{3g}$ &   -
		\\ 
		$A_u$ & $\Delta \sigma_0 s_y \sin k_y + \Delta' \sigma_1(k_x) s_x \sin(k_x/2)$  
		\\ 
		$B_{1u}$ & $\Delta \sigma_0 s_x \sin k_y + \Delta' \sigma_1(k_x)s_y \sin(k_x/2)$  
		\\
		$B_{2u}$ & $\Delta \sigma_2(k_x) s_y \sin(k_x/2) \sin k_y + \Delta' \sigma_1(k_x)s_z \sin(k_x/2)$
		\\ 
		$B_{3u}$ & $\Delta \sigma_0 s_z \sin k_y + \Delta' \sigma_2(k_x) s_x \sin(k_x/2) \sin k_y$
	\end{tabular}
\end{ruledtabular}
	\label{pairpotential}
\end{table}
The energy dispersions in the superconducting states are easily obtained.
For example, at $k_x =\pi$, the BdG Hamiltonian with the $B_{3u}$ pairing is given by
\begin{align}
\nonumber
H(\pi,k_y) = 
&\qty[\qty(c(\pi,k_y) - \mu) \sigma_0 s_0 + \lambda_1 \sin k_y \sigma_3 s_x] \tau_z \\ &+ \qty[\Delta \sin k_y \sigma_0 s_z + \Delta' \sigma_1 s_x \sin k_y] \tau_x, 
\end{align}
and the energy dispersion is
\begin{align}
\nonumber
 E(\pi, k_y) = &\pm \lambda_1 \sin k_y \\ &\pm \sqrt{\qty(\Delta^2 + \Delta'{}^2) \sin^2 k_y + (c(\pi,k_y)-\mu)^2 }.
\end{align}
The energy dispersion has the zeros if the following equation holds
\begin{align}
\sin^2{k_y} = \frac{\qty(m_0 - t_1 + t_2 \cos k_y - \mu)^2}{\lambda^2_1 - \qty(\Delta^2 +\Delta'{}^2)}.
\end{align}
As a result, when $\lambda^2_1 < \Delta^2 +\Delta'{}^2$, there is no node on the $k_x=\pi$ line. 
In the weak-coupling limit ($\Delta, \Delta' \to 0$), on the other hand, the above condition holds and thus superconducting gap nodes appear on the Fermi surface. 
It is consistent with the symmetry consideration in Appendix~\ref{node}. Hereafter, we assume the gapped case $\lambda^2_1 < \Delta^2 + {\Delta'}^2 $, where Majorana Kramers pairs appear.

\subsubsection{Magnetic response}
\label{numerical}

\begin{figure*}
	\includegraphics[width=13.9cm]{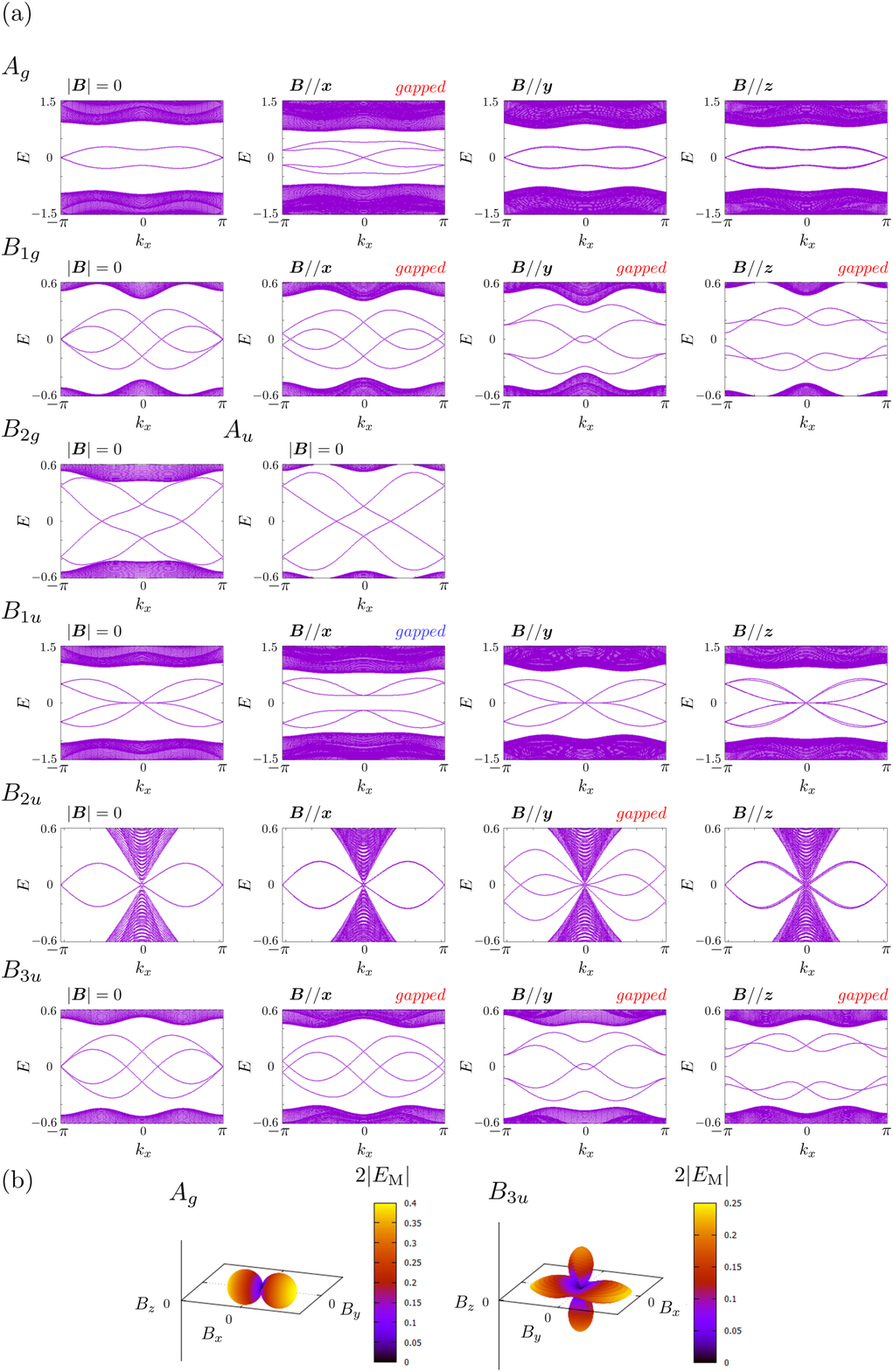}
	\caption{ (a) Energy spectrum for each pair potential in a magnetic field.
	Majorana Kramers pairs with $k_x=\pi$ appear for the $A_g$, $B_{1g}$, $B_{2u}$, and $B_{3u}$ pairings. 
	That with $k_x=0$ appears for the $B_{1u}$ pairing. 
	The $B_{2g}$, $B_{3g}$ and $A_u$ parings have no Majorana Kramers pair. (b) Polar plots of the energy gap $E_{\text{M}}$ of $H+H_{\text{Z}}$ as a function of $\vv{B}$ for the $A_g$ and $B_{3u}$ pairing, respectively. $\mu$ is set to 0 and the other parameters are shown in Table \ref{parameter}.}
	\label{magnetic}
\end{figure*}

We calculate the energy spectrum of the system with $(xz)$ surface in the presence of a Zeeman field.  
The corresponding finite-sized BdG Hamiltonian has the form
\begin{align}
\nonumber
 H(k_x) = &\sum^{N_y}_{n=1}c_n^{\dagger}(k_x)
  \epsilon(k_x)
 c_n(k_x) \\
&+ \sum^{N_y-1}_{n=1} 
\qty[
c_n^{\dagger}(k_x)
 t_y(k_x) 
	c_{n+1}(k_x)
 + \mathrm{h.c.}
 ], 
\label{BdG}
\end{align}
where $N_y$ denotes the number of the sites along the $y$ direction. 
The Fermi energy is set to $0$, then the system has four Fermi surfaces between the $X$ and $S$ points. 
The onsite energy $\epsilon(k_x)$ and hopping $t_y(k_x)$ are obtained by replacing $1 \to c^\dag_n(k_x) c_n(k_x)$,  $\cos k_y \to \frac{1}{2} c^\dag_n(k_x) c_{n+1}(k_x) + \mathrm{h.c.}$, and $\sin k_y \to -\frac{i}{2} c^\dag_n(k_x) c_{n+1}(k_x) + \mathrm{h.c.}$ in the bulk Hamiltonian Eq.~(\ref{BDGD}). 
A magnetic field induces the Zeeman term
\begin{align}
H_{\mathrm Z} = \sum^{N_y}_{n=1} c_n^{\dagger}(k_x) \vv{B} \cdot \vv{s} \sigma_0 \tau_0 c_n(k_x).
\end{align}
The energy spectra of the total Hamiltonians $H(k_x) + H_{\mathrm Z}$ for all the possible pair potentials are shown in Fig.~\ref{magnetic}(a).
Without a magnetic field $|\vv B|=0$, the superconducting gap in the bulk is of the order of $1.0$. 
Majorana Kramers pairs on the surface with $k_x=\pi$ appear for the $A_g$, $B_{1g}$, $B_{2u}$, and $B_{3u}$ pairings, while that with $k_x=0$ appears for the $B_{1u}$ pairing. 
The $B_{3g}$ paring has no Majorana Kramers pair. 
The $B_{2g}$ and $A_u$ pairings also have no Majorana Kramers pair but in-gap surface states protected by glide-plane symmetry (for detail, see Appendix \ref{Z4surface}). 
Figure \ref{magnetic}(b) shows a polar plot of the energy gap $E_{\text{M}}(\vv{B})$ for the $A_g$ pairing with $|\bm{B}|=0.2$. 
The energy gap is of the order of $\sim 0.4$ at maximum for a magnetic field along the $x$ direction. 
Then, the energy gap $E_{\mathrm M}(\boldsymbol{B})$ induced by magnetic fields $\boldsymbol{B}$ is proportional to $E_{\mathrm M}(\boldsymbol{B}) \propto B_x$ for the $A_g$ pairing.
Similarly, the energy gap is proportional to $E_{\mathrm M}(\boldsymbol{B}) \propto B_x$ for the $B_{1u}$ pairing, and $E_{\mathrm M}(\boldsymbol{B}) \propto B_y$ for the $B_{2u}$ pairing, respectively. 
This Majorana-Ising-spin nature is consistent with that predicted by the general theory shown in Table \ref{Pmma}.
The magnetic response for the $B_{1g}$ pairing is the same as for $B_{3u}$, which has been discussed in Sec.~\ref{General classification}. 
Fig.~\ref{magnetic}(b) shows a polar plot of the energy gap $E_{\text{M}}(\vv{B})$ for the $B_{3u}$ pairing with $|\bm{B}|=0.2$. 
The energy gap is of the order of $\sim 0.25$ at maximum and biaxially (quadrupolar) anisotropic, which is given by $E_{\text{M}}(\boldsymbol{B}) \sim \sqrt{\rho_{xx}B^2_x + \rho_{yy}B^2_y + \rho_{zz}B^2_z} - \sqrt{\rho'_{zz} B^2_z}$. 
Coefficients $\rho_{ii}$ and $\rho_{ii}'$ are material parameters.

\section{Superconducting gap and node for space group \texorpdfstring{\it{Pmma}}{Pmma}}
\label{node}

The symmetry-allowed gap functions are classified based on the group theoretical scheme \cite{Kobayashi180504, Sumita134512, Sumita134513, Yoshida235105}.
Let $M_{\vv k} = G_{\boldsymbol{k}} + G_{\boldsymbol{k}} I\Theta$ be a little group mapping the momentum as $\vv k \rightarrow g\vv k = \vv k$ for the high symmetry line. 
$G_{\boldsymbol{k}}$ is the little group of space group $G$, and $I\Theta$ denotes space-time inversion.
The character $\chi[P_{\boldsymbol{k}}(m)]$ for the representation $P_{\boldsymbol{k}}$ of the Cooper pair formed by $\gamma_{\boldsymbol{k}}$ and $\gamma_{-\boldsymbol{k}}$, which are irreps of $M_{\boldsymbol{k}}$ and $M_{-\boldsymbol{k}}$, is obtained by using the Mackey-Bradley theorem:  
\begin{align}
\chi\qty[P_{\vv k}(m)] &= \chi\qty[\gamma_{\vv k}(m)]\chi\qty[\gamma_{\vv k}(ImI)], 
\label{MB}
 \\
\chi\qty[P_{\vv k}(Im)] &= -\chi\qty[\gamma_{\vv k}(ImIm)].
\label{MB2}
\end{align}
The symmetry operations $g = \qty{R_g|\boldsymbol{\tau}_g}$, $Ig = \qty{I R_g|-\boldsymbol{\tau}_g}$, $IgI=\qty{R_g|-\boldsymbol{\tau}_g}$, and $IgIg = \qty{R_g^2|R_g\boldsymbol{\tau}_g - \boldsymbol{\tau}_g}$ are listed in Table \ref{g_Pmma}.
\begin{table}
	\caption{Symmetry operation $g$ of $Pmma$. $I$ denotes the spatial inversion. The superscript $^d$ is indicated for the $2\pi$ rotation.}
	\begin{ruledtabular}
    \begin{tabular}{llll}
    	$g$ & $Ig$ & $IgI$ & $IgIg$
    	\\
    	\hline
    	$\qty{E|\boldsymbol{0}}$ & $\qty{I|\boldsymbol{0}}$ & $\qty{E|\boldsymbol{0}}$ & $\qty{E|\boldsymbol{0}}$
    	\\
    	$\qty{C_2(z)|\boldsymbol{a}/2}$ & $\qty{\sigma(xy)|-\boldsymbol{a}/2}$ & $\qty{C_2(z)|-\boldsymbol{a}/2}$ & $\qty{{}^dE|-\boldsymbol{a}}$
    	\\
    	$\qty{C_2(y)|\boldsymbol{0}}$ & $\qty{\sigma(xz)|\boldsymbol{0}}$ & $\qty{C_2(y)|\boldsymbol{0}}$ & $\qty{{}^dE|\boldsymbol{0}}$
    	\\
    	$\qty{C_2(x)|\boldsymbol{a}/2}$ & $\qty{\sigma(yz)|-\boldsymbol{a}/2}$ & $\qty{C_2(x)|-\boldsymbol{a}/2}$ & $\qty{{}^dE|\boldsymbol{0}}$
    	\\
    	$\qty{I|\boldsymbol{0}}$ & $\qty{E|\boldsymbol{0}}$ & $\qty{I|\boldsymbol{0}}$ & $\qty{E|\boldsymbol{0}}$
    	\\
    	$\qty{\sigma(xy)|\boldsymbol{a}/2}$ & $\qty{C_2(z)|-\boldsymbol{a}/2}$ & $\qty{\sigma(xy)|-\boldsymbol{a}/2}$ & $\qty{{}^dE|\boldsymbol{0}}$
    	\\
    	$\qty{\sigma(xz)|\boldsymbol{0}}$ & $\qty{C_2(y)|\boldsymbol{0}}$ & $\qty{\sigma(xz)|\boldsymbol{0}}$ & $\qty{{}^dE|\boldsymbol{0}}$
    	\\
    	$\qty{\sigma(yz)|\boldsymbol{a}/2}$ & $\qty{C_2(x)|-\boldsymbol{a}/2}$ & $\qty{\sigma(yz)|-\boldsymbol{a}/2}$ & $\qty{{}^dE|-\boldsymbol{a}}$
    \end{tabular}
	\end{ruledtabular}
\label{g_Pmma}
\end{table}
The result is summarized in Table \ref{gap node}. 
The irreducible decomposition of $P_{\boldsymbol k}$ shows symmetry-adapted gap functions. 
Pair potentials that are not included in the decomposition cannot create the superconducting gap, resulting in gap nodes. 
The details of the calculation are shown in the following subsections. 
\begin{table*}
	\caption{Classification of gap functions in space group $Pmma$. The characters for representations $P_{\boldsymbol{k}}$ of the pair potential and the irreducible decomposition of possible pair potential are shown. }
	\begin{ruledtabular}
	\begin{tabular}{ccccccccccc}
	Line & $(k_x,k_y)$ & $E$ & $C_2(z)$ & $C_2(y)$ & $C_2(x)$ & $I$ & $\sigma(xy)$ &$\sigma(xz)$ & $\sigma(yz)$ & Decomposition
		\\
		\hline
	$\Lambda$ & $(0,0)$ & $4$ & $0$ & $2$ & $2$ & $-2$ & $2$ & $0$ & $0$ & $A_g + A_u + B_{2u} + B_{3u}$
		\\
	$G$ &	$(\pi,0)$ & $16$ & $0$ & $4$ & $-4$ & $-4$ & $-4$ & $0$ & $0$ & $A_g + B_{1g}  + 3B_{2g} + B_{3g} + 3A_u + 3B_{1u} + 3B_{2u} + B_{3u}$
		\\
	$H$ &	$(0,\pi)$ & $4$ & $0$ & $2$ & $2$ & $-2$ & $2$ & $0$ & $0$ & $A_g + A_u + B_{2u} + B_{3u}$
		\\
	$Q$ &	$(\pi,\pi)$ & $16$ & $0$ & $4$ & $-4$ & $-4$ & $-4$ & $0$ & $0$ & $A_g + B_{1g}  + 3B_{2g} + B_{3g} + 3A_u + 3B_{1u} + 3B_{2u} + B_{3u}$
		\\
		\hline\hline
	Line & $(k_y,k_z)$ & $E$ & $C_2(z)$ & $C_2(y)$ & $C_2(x)$ & $I$ & $\sigma(xy)$ &$\sigma(xz)$ & $\sigma(yz)$ & Decomposition
		\\
		\hline
	$\Sigma$ & $(0,0)$ & $4$ & $2$ & $2$ & $0$ & $-2$ & $0$ & $0$ & $2$ & $A_g + A_u + B_{1u} + B_{2u}$
		\\
	$C$ &	$(\pi,0)$ & $4$ & $2$ & $2$ & $0$ & $-2$ & $0$ & $0$ & $2$ & $A_g + A_u + B_{1u} + B_{2u}$
		\\
	$A$ &	$(0,\pi)$ & $4$ & $2$ & $2$ & $0$ & $-2$ & $0$ & $0$ & $2$ & $A_g + A_u + B_{1u} + B_{2u}$
		\\
	$E$ &	$(\pi,\pi)$ & $4$ & $2$ & $2$ & $0$ & $-2$ & $0$ & $0$ & $2$ & $A_g + A_u + B_{1u} + B_{2u}$
		\\
		\hline\hline
	Line &	$(k_z,k_x)$ & $E$ & $C_2(z)$ & $C_2(y)$ & $C_2(x)$ & $I$ & $\sigma(xy)$ &$\sigma(xz)$ & $\sigma(yz)$ & Decomposition
		\\
		\hline
	$\Delta$ &	$(0,0)$ & $4$ & $2$ & $0$ & $2$ & $-2$ & $0$ & $2$ & $0$ & $A_g + A_u + B_{1u} + B_{3u}$
		\\
	$B$ &	$(\pi,0)$ & $4$ & $2$ & $0$ & $2$ & $-2$ & $0$ & $2$ & $0$ & $A_g + A_u + B_{1u} + B_{3u}$
		\\
	$D$ &	$(0,\pi)$ & $4$ & $2$ & $0$ & $-2$ & $-2$ & $0$ & $2$ & $4$ & $A_g + 2B_{1u} + B_{2u}$
		\\
	$Q$ &	$(\pi,\pi)$ & $4$ & $2$ & $0$ & $-2$ & $-2$ & $0$ & $2$ & $4$ & $A_g + 2B_{1u} + B_{2u}$
	\end{tabular}
	\end{ruledtabular}
	\label{gap node}
\end{table*}

\begin{figure*}
	\includegraphics[width=13cm]{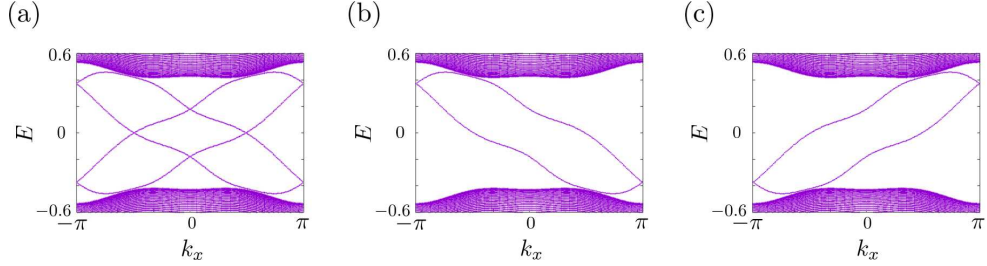}
	\caption{Energy spectrum for the $B_{2g}$ pairing of (a) $H(k_x)$ (the same as in Fig. \ref{magnetic}), (b) $H_+(k_x)$, and (c) $H_-(k_x)$. 
		For the in-gap surface states, only those localized at the $n = 1$ edge, which satisfy Eq. (\ref{leftedge}), are shown. 
	}
	\label{Z4}
\end{figure*}
\subsection{Classification on the \texorpdfstring{$\Lambda$}{Lambda} line \texorpdfstring{$(k_x,k_y,k_z)=(0,0,k_z)$}{(kx, ky, kz) = (0, 0, kz)}}

On the $\Lambda$ line $k_x = k_y = 0$, the little group is written as
\begin{align}
G_{\vv k} &= \{E|\vv{0} \}T + \{C_2(z)|\vv{a}/2 \}T + \{\sigma(xz)|\vv{0} \}T 
\nonumber\\ & \quad
+ \{\sigma(yz)|\vv{a}/2 \}T,
\label{G}
\end{align}
where $T$ is the translation subgroup. 
There is the two-dimensional double-valued pseudoreal irrep $\Lambda_5$ of $G_{\boldsymbol{k}}$~\cite{Elcoro}, which is time-reversal-invariant. 
The character is zero except for $m=E$; 
$\chi[\gamma_{\boldsymbol{k}}(E)] = 2$ and
$\chi[\gamma_{\boldsymbol{k}}(m \ne E)] = 0$. 
Substituting these into Eqs. (\ref{MB}) and (\ref{MB2}), one obtains $\chi[P_{\boldsymbol{k}}(m)]$.
Note that representations for the Cooper pair are translation invariant, i.e., $P_{\boldsymbol{k}}(m) = P_{\boldsymbol{k}}(m t)$ for $t \in T$.
We obtain $\chi[P_{\vv k}(m)]$ for the $H$ [$(k_x, k_y) = (0, \pi)$], $\Sigma$ [$(k_y, k_z) = (0, 0)$], and $\Delta$ [$(k_z, k_x) = (0, 0)$] lines in a similar way. 
In the following, we focus on the situations that $\chi[P_{\vv k}(m)]$ is different from the above.

\subsection{Classification on the G line \texorpdfstring{$(k_x, k_y, k_z) = (\pi, 0, k_z)$}{(kx, ky, kz) = (pi, 0, kz)} and the Q line \texorpdfstring{$(\pi, \pi, k_z)$}{(pi, pi, kz)}}

On the $G$ ($Q$) line, there is the two-dimensional double-valued real irrep $G_5$ ($Q_5$), whose characters are zero except for the identity. 
The time-reversal-invariant irrep is given by two copies of $G_5$ ($Q_5$). Thus the characters are $\chi[\gamma_{\boldsymbol{k}}(E)]=4$ and $\chi[\gamma_{\boldsymbol{k}}(m \ne 0)]=0$.
By using the Mackey-Bradley theorem [Eqs.~(\ref{MB}) and (\ref{MB2})], one obtains the character $\chi[P_{\boldsymbol{k}}(m)]$.

\subsection{Classification on the D line \texorpdfstring{$(k_z,k_x,k_y)=(0,\pi,k_y)$}{(kz, kx, ky) = (0, pi, ky)} and the P line \texorpdfstring{$(\pi,\pi,k_y)$}{(pi, pi, ky)}}

On the $D$ and $P$ lines, $G_{\vv k}$ is written as
\begin{align}
G_{\vv k} &= \{E|\vv{0} \}T + \{C_2(y)|\vv{0} \}T + \{\sigma(xy)|\vv{a}/2 \}T 
\nonumber\\ & \quad
+ \{\sigma(yz)|\vv{a}/2 \}T.
\end{align}
There is a pair of the double-valued conjugated irreps $(D_4, D_5)$. 
The time-reversal-invariant irreps that are the direct sum of these conjugated representations are given by
\begin{align}
\gamma_{\vv k}(\{E| \vv 0\}) &= \pmqty{
1 & 0 \\
0 & 1
}, 
\\
\gamma_{\vv k}(\{C_2(y)| \vv 0\}) 
&= \pmqty{
-i & 0 \\
0 & i
}, 
\\
\gamma_{\vv k}(\{\sigma(xy)| \vv{a}/2\}) 
&= \pmqty{
1 & 0 \\
0 & -1
},
\\
\gamma_{\vv k}(\{\sigma(yz)| \vv{a}/2\}) 
&= \pmqty{
i & 0 \\
0 & i
}.
\end{align}
By using the Mackey-Bradley theorem [Eqs.~(\ref{MB}) and (\ref{MB2})], the character of $P_{\vv k}(m)$ is calculated.

\section{Surface states protected by the \texorpdfstring{$\mathbb Z_4$}{Z4} invariant}
\label{Z4surface}

From Fig.~\ref{magnetic}, the $B_{2g}$ and $A_u$ pairings host two right/left-moving surface states within the superconducting gap instead of Majorana Kramers pairs on the zero energy.
These in-gap states are protected by the $\mathbb Z_4$ invariant from the glide-plane symmetry, as discussed below. 
 
The BdG Hamiltonian is decomposed into those in the eigenspaces of glide plane $\tilde{D}_{\bm k_{\parallel}}(U_2 = \qty{\sigma(xy) | \bm a/2})$, 
i.e., 
$H(k_x) = H_+(k_x)  \oplus H_-(k_x) $, where the subscript $\pm$ denotes the eigenvalue $\pm i e^{i k_x/2}$. 
The Hamiltonian $H_{\pm}(k_x)$ has the same form as Eq.~(\ref{BdG}). 
We define the eigenstates of $H_{\pm}(k_x)$ by 
\begin{align}
 H_{\pm}(k_x)|k_x, \alpha \rangle_{\pm} = E^{\pm}_n(k_x)|k_x, \alpha \rangle_{\pm}.
\end{align}
There are two surface states $|k_x,\alpha \rangle_{\pm} \ (\alpha=1,2,3,4)$ for any $k_x$ bound around each surface $n \sim 1$ ($n \sim N_y$). 
In Fig.~\ref{Z4}, we plot the energy spectrum for the eigenstates localized on $n \sim 1$ that satisfy
\begin{align}
\nonumber
\sum^{N_y/4}_{n=1} 
|\langle 0 | c_{\pm,n}(k_x) &|k_x,\alpha \rangle_{\pm}|^2 \\ &> \sum^{N_y}_{n=3N_y/4} |\langle 0 | c_{\pm,n}(k_x) |k_x,\alpha \rangle_{\pm}|^2,
\label{leftedge}
\end{align}
for $\alpha=1,2,3$, and 4, where $|0 \rangle$ denotes the vacuum and $c_{\pm,n}(k_x)$ denotes the electron operator on each sector. 

The above condition means that the number of quasiparticles existing in $1 \leq n \leq N_y/4$ is larger than that in $3N_y/4 < n < N_y$. 
We can see the M\"{o}bius structure with $4\pi$ periodicity of the surface states shown in Figs.~\ref{Z4}(b) and \ref{Z4}(c), protected by the $\mathbb Z_4$ invariant of $\theta=2$ \cite{Shiozaki195413, Yoshida235105, Daido227001}.

\bibliography{ref}

\begin{thebibliography}{106}
\bibitem [{mul()}]{multi}%
  \BibitemOpen
  \href@noop {} {}\bibinfo {note} {A multidimensional irreducible
  representation is regarded as one-dimensional irreducible representations of
  the subgroup.}\BibitemShut {Stop}%
\bibitem [{UPt()}]{UPt3}%
  \BibitemOpen
  \href@noop {} {}\bibinfo {note} {The recent experiment reported that the
  polar Kerr effect appears below the lower superconducting transition
  temperature at zero fields, which implies the spontaneous breaking of TRS in
  the lower-temperature superconducting phase \cite{Schemm190}. According to
  this observation, it appears that the chiral $f$-wave pairing in the $E_{1u}$
  or $E_{2u}$ state is rather plausible. Therefore, the absence (presence) of
  anisotropic magnetic response might signal TRS-breaking (preserving) pairing
  in UPt$_3$.}\BibitemShut {Stop}%
\end{thebibliography}
\end{document}